\documentclass[twocolumn]{aa}
\usepackage{graphicx}
\newcommand{\galics}{{\sl GalICS}}
\newcommand{\momaf}{{\sl MoMaF}}

\def\hmpc{\rm \,h^{-1}\,Mpc}

\def\xir{$\xi(r)$\ }
\def\xis{$\xi(s)$\ }
\def\xip{$\xi(r_p,\pi)$\ }
\def\xip{$\xi(r_p,\pi)$\ }
\def\wp{$w_p(r_p)$\ }

\def\n_med{{\left<n\right>}}

\def\begc{\begin{center} }
\def\endc{\end{center} }
\def\begf{\begin{figure} }
\def\endf{\end{figure} }
\def\j3{{J_3}}
\begin{document}
   \title{The VIMOS VLT Deep Survey
         \thanks{based on data
         obtained with the European Southern Observatory Very Large
         Telescope, Paranal, Chile, program 070.A-9007(A), and on data
         obtained at the Canada-France-Hawaii Telescope, operated by
         the CNRS of France, CNRC in Canada and the University of Hawaii}
}

   \subtitle{Computing the two
        point correlation statistics and associated uncertainties}

   \author{A.~Pollo \inst{1},
           B.~Meneux \inst{2},
           L.~Guzzo  \inst{1},
           O.~Le F\`evre \inst{2},
J. Blaizot \inst{2},
A. Cappi \inst{3}, 
A. Iovino \inst{4},
C. Marinoni \inst{2},
H.J. McCracken \inst{5},\inst{6}
D. Bottini \inst{7}, 
B. Garilli \inst{7}, 
V. Le Brun \inst{2}, 
D. Maccagni \inst{7}, 
J.P. Picat \inst{8},  
R. Scaramella \inst{9}, 
M. Scodeggio \inst{7}, 
L. Tresse  \inst{2},
G. Vettolani \inst{10},
A. Zanichelli \inst{10}, 
C. Adami \inst{2}, 
M. Arnaboldi \inst{13},
S. Arnouts \inst{2},
S. Bardelli \inst{3},
M. Bolzonella \inst{3},
S. Charlot \inst{5},\inst{11},
P. Ciliegi \inst{10},
%(moved)
T. Contini \inst{8},
S. Foucaud \inst{7},  
P. Franzetti \inst{7},
I. Gavignaud \inst{8}, 
O. Ilbert \inst{2}, 
B. Marano \inst{12}, 
G. Mathez \inst{8},
A. Mazure \inst{2},
R. Merighi \inst{3}, 
S. Paltani \inst{2},
R. Pell\`o \inst{8}, 
L. Pozzetti \inst{3},
M. Radovich \inst{13}, 
G. Zamorani  \inst{3}, 
E. Zucca  \inst{3}
M. Bondi \inst{10}, 
A. Bongiorno \inst{3}
G. Busarello \inst{13}, 
L. Gregorini \inst{10},
Y. Mellier \inst{5},\inst{6} 
P. Merluzzi \inst{13}, 
V. Ripepi \inst{13},
D. Rizzo \inst{8} 
          }

   \offprints{A. Pollo}

   \institute{
        INAF - Osservatorio Astronomico di Brera, via Bianchi, 46, Merate (LC), 
Italy \\
        e-mail: apollo@merate.mi.astro.it
        \and
Laboratoire d'Astrophysique de Marseille, UMR 6110 CNRS-Universit\'e
de Provence, Traverse du Siphon-Les trois Lucs, 13012 Marseille, France
\and
INAF - Osservatorio Astronomico di Bologna, via Ranzani 1, 40127 Bologna, Italy
\and
INAF - Osservatorio Astronomico di Brera, via Brera, Milan, Italy
\and
Institut d'Astrophysique de Paris, UMR 7095, 98 bis Bvd Arago, 75014 Paris, France
\and
Observatoire de Paris, LERMA, UMR 8112, 61 Av. de l'Observatoire, 75014 Paris, France
\and
IASF - INAF, Milano, Italy
\and
Laboratoire d'Astrophysique - Observatoire Midi-Pyr\'en\'ees, Toulouse, France
\and
INAF - Osservatorio Astronomico di Roma, Italy
\and
INAF - Istituto di Radio-Astronomia, Bologna, Italy
\and
Max Planck Institut fur Astrophysik, 85741 Garching, Germany
\and 
Universit\`a di Bologna, Departimento di Astronomia, via Ranzani 1, 40127 Bologna, Italy
\and
INAF - Osservatorio Astronomico di Capodimonte, via Moiariello 16, 80131 Napoli, Italy
             }

   \date{Received September 7, 2004; accepted March 10, 2005}

    \abstract{We present
% are presenting in this paper 
a detailed description
%account 
of
      the methods used
      to compute the three-dimensional two-point galaxy correlation
      function in the VIMOS-VLT deep survey (VVDS).
      We investigate how
      instrumental selection effects and observational biases affect
      the measurements and identify the methods to correct for them.  We
      quantify the accuracy of our
% correction method 
       corrections using an
      ensemble of  $50$ mock galaxy surveys generated with the \galics{}
      semi-analytic model of galaxy formation which
      incorporate the %same 
      selection biases and tiling strategy 
% as 
      of the real data. 
%does
We demonstrate that we are able to recover the 
      real-space two-point correlation function $\xi(s)$ and the projected correlation function $w_p(r_p)$ to an accuracy better than 10\%
      on scales larger than 1 h$^{-1}$ Mpc
% , and of
%      about 30\% on scales below 1 h$^{-1}$ Mpc,
      with the sampling strategy used for the first epoch
      VVDS data. 
%The projected correlation function $w_p(r_p)$ 
%      is recovered with an accuracy better than 10\%
%      on all scales $0.1 \leq r \leq 10$ h$^{-1}$ Mpc.
%      There is a tendency for a small
%      but systematic 
%{\it under}
%      {\bf over}-estimate of the correlation
%      length derived from $w_p(r_p)$ of 
%
%      {\bf 5}\% on average,
%      remaining after our correction process. 
      The large number of simulated surveys allows us to provide a reliable
      estimate of the cosmic variance on the measurements of 
      the correlation length $r_0$ at $z \sim 1$, of about 15--20\%
      for the first epoch VVDS observation 
%(\cite{LeFevre2004}) 
while any residual systematic effect in the measurements of $r_0$ is always
      below $5\%$. The
      error estimation and measurement techniques outlined in this paper
      are being used in several parallel studies which investigate in
      detail the clustering properties of galaxies in the VVDS.

   \keywords{cosmology: deep redshift surveys -- 
             large scale structure of Universe --
             methods: statistical -- galaxies: evolution
               }
   }
\authorrunning{A. Pollo et al.}
\titlerunning{VVDS Correlation Statistics}
    \maketitle
%________________________________________________________________

\section{Introduction}

The VIMOS VLT Deep Survey (VVDS, \cite{LeFevre2004}) is dedicated to
study the evolution of galaxies and large scale structure to $z \sim 2$
with a significant fraction of galaxies reaching $z \sim 4 $.  The VVDS
spectroscopic survey is performed with the VIMOS spectrograph at the European
Southern Observatory Very Large Telescope
% (\cite{VIM})
and
complemented with multi-color BVRI imaging
data obtained at the CFHT telescope (\cite{hjmcc}, \cite{lefevre04}).
% and deep spectroscopy, obtained 
The complete survey will consist of four fields of $2^\circ$ by
$2^\circ$ each, with multi-band photometry coverage in the BVRI (and
partly UJK) bands.  Multi-object spectroscopy down to $I_{AB} = 22.5$
is being obtained over the four fields (``VVDS Wide''), with a deeper
area of $1.5$ deg$^2$ in the VVDS-02h and in the Chandra Deep Field
South (VVDS-CDFS) covered to $I_{AB} = 24$ (``VVDS Deep'').  The first
epoch VVDS data consist of more than 11000 spectra obtained in the
VVDS-Deep fields (\cite{LeFevre2004}).

One of the key science goals of the VVDS is to measure the evolution of
galaxy clustering from the present epoch up to $z \sim 2$.  The
simplest statistic used for this analysis is the spatial two-point 
correlation function $\xi(r)$ and its variants,
(e.g.~\cite{peebles80}), i.e. the second moment of the galaxy
distribution.  Given the geometry and selection function of galaxy
surveys, however, the practical estimation of $\xi(r)$ from the actual
data is not straightforward. Edge effects, sampling inhomogeneities and
selection effects all introduce different biases that hamper the
survey's ability to estimate the true underlying clustering process.
Moreover, intrinsic systematic uncertainties due to the limited size of
the volume of the Universe explored (``cosmic variance'') need to be
accounted for when computing realistic error bars on the measured
correlation values. 
%The VVDS is no exception in this respect, and the
%computation of two-point correlation functions from these data requires
%a proper treatment of all these effects.  In this paper, we are discussing
%in detail the biases specific to the VVDS and the methods that have been
%devised for their correction.  Through realistic mock samples,
%constructed from state-of-the-art simulations and reproducing in detail
%the true survey, we are providing a direct
%test of the accuracy of our corrections and we 
%determine the overall errors
%expected for the measured quantities.

The aim of this paper is to present a comprehensive description 
of the biases specific to the VVDS, along with the methods we developed 
to correct for them. The strategy we
adopt relies on the construction of realistic ``pre-observation'' mock
catalogs using the \momaf{} software (\cite{gal2}) and the \galics{}
hybrid model for galaxy formation (\cite{gal1}). We then observe these
mock catalogs, by mimicking the relevant observational
selections and biases. Comparing original and observed mock surveys 
allows us to (i) quantitatively understand  the impact of the different biases
inherent to the VVDS data on clustering estimates, and (ii) to explore
and validate methods that allow us to recover the original signal.   This
strategy is possible because \galics{} predictions have been shown to
agree fairly well with a wide range of observations (e.g.
\cite{gal1}, \cite{gal3}), and is thus expected to yield catalogs
realistic enough to carry out a convincing consistency check. Because
our mock catalogs contain realistic clustering properties, we can
also use them to predict the cosmic variance amplitude in order to compute
realistic errors on the clustering estimates we will perform on the
real data.

The paper is organized as follows. In section 2 we discuss the
different kind of biases expected in the current VVDS first-epoch data.
In section 3 we discuss the construction of mock VVDS catalogs from the
\galics/{\sl MoMaf} simulations which assume a flat Cold Dark Matter model
with $\Omega_m = 0.333$, $\Omega_{\Lambda} = 0.667$ and $h = 0.667$. 
In section 4 we present the definitions of the 
two-point correlation functions. Then, in section 5 we discuss the details of 
the error measurement strategy when applied
to VVDS. In section 6 we show how the measured
two-point correlation function is affected by the features particular
to our survey and we discuss the methods developed to
correct for these biases and properly estimate the correlation function
$\xi(r_p,\pi)$, its projection $w_p(r_p)$, and the correlation length
$r_0$ and slope $\gamma$, as a function of redshift.  
Section 7 summarizes our results.

\section{The selection function of VVDS first epoch observations}

The first epoch spectra of the VVDS-Deep 
collected during the 2002 and 2003 campaigns
are concentrated within the 02h deep field, and the CDFS
(\cite{LeFevre2004}).  
First epoch spectra have been collected for galaxies down to $I_{AB}
\leq 24$ in the  
%$1.3^\circ \times1^\circ$ 
$0.61$ sq. degree
sub-area of the VVDS-02h field and a region of $21\times21.6$ sq.
arcminutes centered on the Chandra Deep Field South (CDFS,
\cite{Giacconi02}).  
The VVDS First Epoch data geometrical lay-out,
sampling rate and incompleteness are used
as a reference benchmark in this paper.
        
\subsection{Catalog structure and biases}

%______________________________________________
   \begin{figure}
   \centering
   \includegraphics[width=8.5cm]{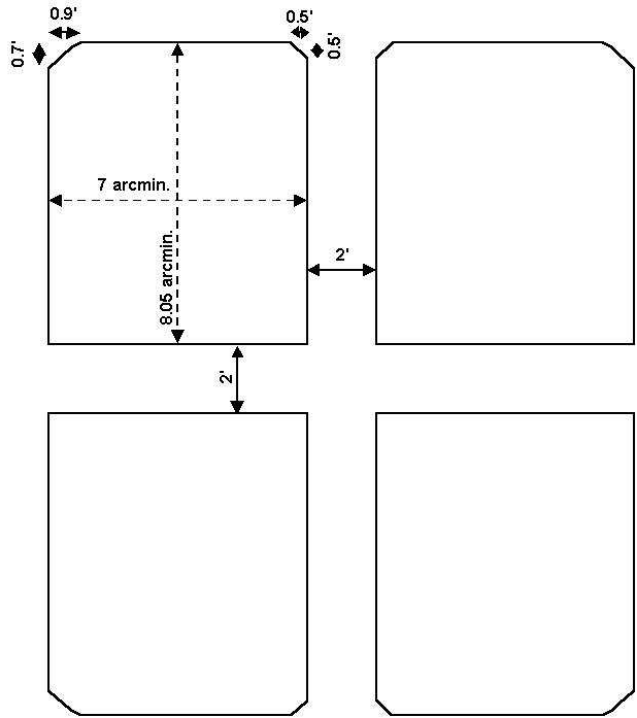}
   \caption{Lay-out of the VIMOS field of view.   INVAR masks with laser-cut 
        slits are placed on the focal plane within the four rectangular areas 
        (``VIMOS channels''). 
        }
              \label{VIMex}
    \end{figure}

  \begin{figure*} \centering
  \includegraphics[width=8.5cm]{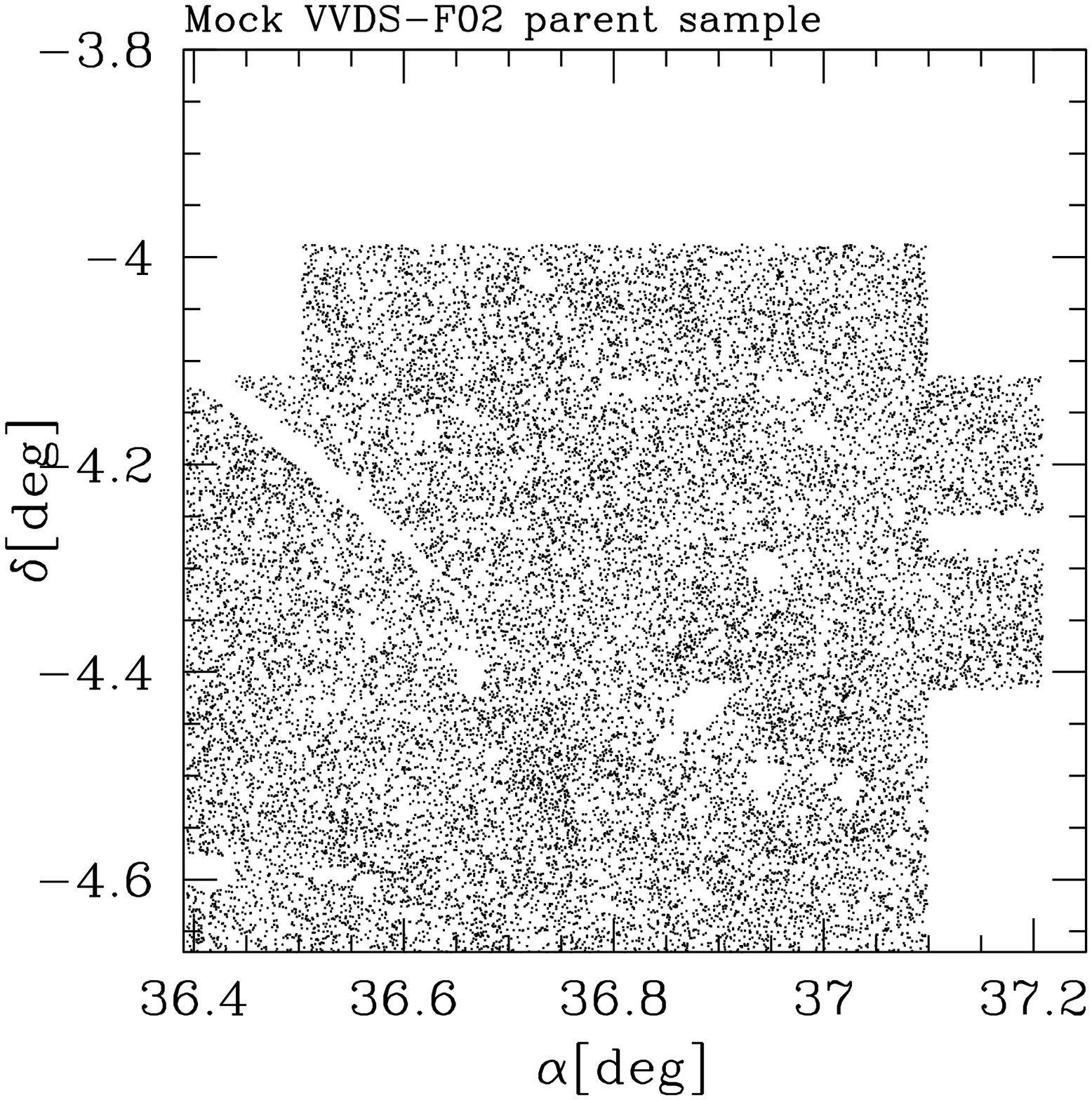}
  \includegraphics[width=8.5cm]{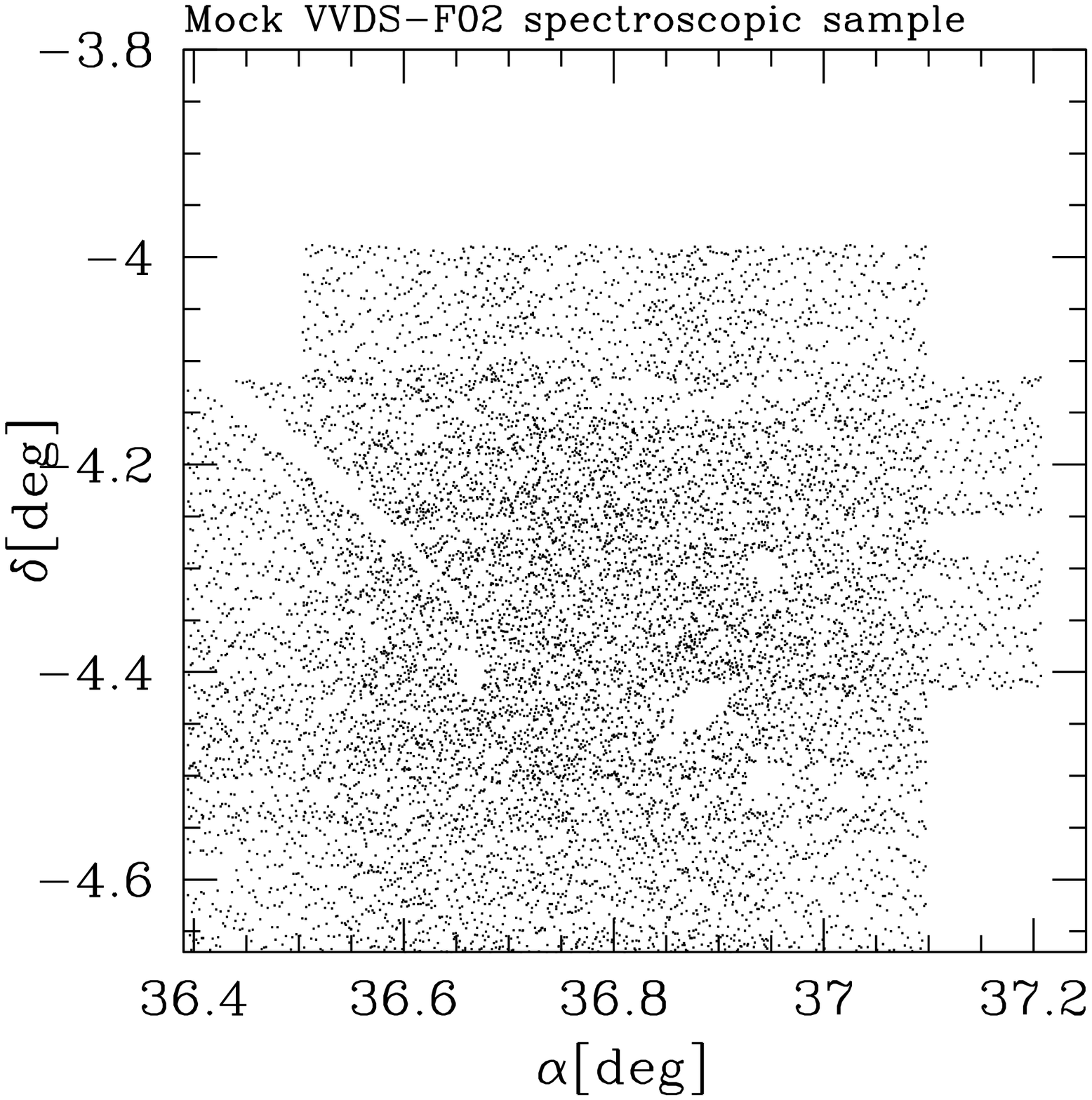} \caption{Galaxy
  distribution in a mock VVDS-02h catalog, constructed using the
  \galics{} simulations with the same lay-out as the 20 observed
  pointings in the actual first-epoch VVDS field and applying the full
  range of selection effects present in the data, as e.g. the
  photometric mask.  The left panel shows the parent photometric
field, including all objects with $I_{AB} \leq 24$ within the current
VVDS-02h boundaries and mask . In the right panel only the objects
selected for spectroscopy are shown. 
  %LG Note the complex density patterns in the (right) field,
%in the selected spectroscopic targets positions.  
Note the density gradient towards the central part of the field,
due to multiple passes over the same area.
% is evident.
        }
              \label{VIRcomp}
    \end{figure*}
 
A number of factors, both in the parent photometric catalog from
which the target galaxies are selected and in the way the
spectroscopic observations are carried out, contribute to create
selection effects that bias any estimate of galaxy clustering
if 
%no care is taken to correct for them:
not properly accounted for.

\begin{figure}
   \centering
   \includegraphics[width=8.5cm]{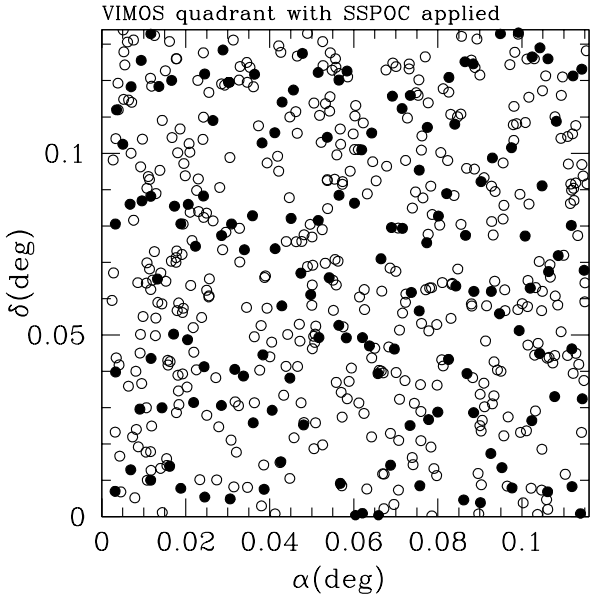}
   \caption{
%{\it R! caption changed to make it shorter and more clear}
Spectroscopic targets (filled circles) selected in one of the four VIMOS
quadrants from a complete VVDS mock 
%catalog made from \galics{}simulations.  
%The parent 
photometric sample  (open circles). 
%{\it R! All the galaxies present in the field in }
%covers the magnitude range $I_{AB} = 17.5 - 24$ mag as in the real case.
%{\it R! (chosen so for clarity) are represented by open circles, while
%objects chosen for spectroscopy are shown as full circles and }
Note how the optimization software tends to select spectroscopic
targets aligned along horizontal rows, while, clearly, very close
pairs are not observed. Typically, however, 4 independent
observations are conducted on the same area, each with a similar
target layout, but shifted by a few arcminutes.  This significantly reduces 
both the alignment and proximity effects.  The residual bias is then further
corrected by the weighting scheme discussed in \S~4.  Overall, the
four passes produce a typical sampling rate of one galaxy in four.
%the underlying magnitude selected population of $\simeq25$\% 
} 
\label{SSfield} 
\end{figure}

\begin{enumerate} \item {\bf Photometric defects.}  Some areas are
excised from the I-band CCD images during their photometric analysis,
due 
%in particular 
to the presence of bright stars or other instrumental
effects (e.g. stray-light from a bright star outside the field of
view).  The resulting photometric galaxy catalog, therefore, features
some artificially empty regions.
        
\item {\bf VIMOS lay-out.}  The field of view of the VIMOS spectrograph
  consists of four $7'$ by $8'$ quadrants, separated by $2'$ gaps, as
  shown schematically in Figure~\ref{VIMex}.  At the typical resolution
  used in the VVDS, between 110 and 150 spectra are collected in each
  quadrant during a single observation.  Clearly, no galaxies are
  observed over the area of the ``cross'' between the four quadrants,
  unless one observes the area with a new pointing, shifted with respect to the
  first one (see below).
        
\item {\bf Missing quadrants.} For a few pointings, one or two
  quadrants can be ``blind'', i.e. with no spectra observed due to a
  misplacement of the multi-slit masks during the observations.
      
\item {\bf Incomplete coverage.}  The planned final area is being
  covered through a mosaic of adjacent pointings.  Thus, at any
  intermediate stage the available spectral data set is distributed in
  a non-uniform fashion on the sky.  The largest contiguous area
  currently covered in the 02h deep field corresponds to about $0.5$
  square degrees, with the geometry shown in Figure~\ref{VIRcomp}.
  
\item {\bf Varying sampling density.} The VVDS observational strategy
  involves multiple passes over the same area to increase the
  spectral sampling rate.  While a central region of the 02h deep field
  is exposed 4 times (i.e. it is visited by four independent pointings
  with different slit masks), the external areas are covered only twice
  due to the tiling strategy. During subsequent observing runs, the
  VIMOS pointings are shifted with respect to the previous ones 
%by a random amount, 
  usually by around $2'$,  to ensure that the cross visible
  in Figure~\ref{VIMex} is filled. As a consequence, the mean surface
  density of 
observed
objects 
%with observed spectra 
varies across the field.
        
\item {\bf Optimization of the number of slits and mechanical
    constraints.} A specific source of bias in the VIMOS observations
  is introduced by VMMPS - the VIMOS Mask Manufacturing Preparation
  Software, and specifically by the Super-SPOC code (\cite{SSPOC}).
  The width
  of a slit is set to 1 arcsecond (or about 5 detector pixels),
  and its typical length 
%of a slit 
is $\sim6-10$ arcseconds to
  include both the object of interest and enough information on the
  sky spectral background to correct for it. 
  The VMMPS software automatically allocates slits to objects in the input
  catalog with the goal of maximizing the total number of spectra. In
  general, this means that the spectroscopic sample is not a random
  sparse sampling of the clustering pattern over the sky, but a
more homogeneous sub-sample.
  Specifically, VMMPS tends to place objects in rows, so to
maximize the number of spectra across the CCD (see
Figure~\ref{SSfield}), with an additional
%{\it R! In addition, It has }
slight preference towards objects of small angular size.  
%{\it R! Finally, due to the typical size of slits, the fixed direction of the
%slits placed East-West,   and the fixed  space occupied on the CCD by the spectrum, }
As typical with multi-object spectrographs, the minimum slit size implies that, after one single
spectroscopic pass, there is a bias against observing very close
angular pairs on the sky. Having multiple passes, however, significantly improves
 the situation, allowing for very close pairs to be
observed in subsequent exposures.

\end{enumerate}

The final spectroscopic sample is thus affected to different degrees
by all these factors.  Figure~\ref{VIRcomp} shows the current lay-out
of the observed pointings in the 02h field, compared to the parent
photometric sample over the same area. Features from the two
main effects are 
%evident
obvious from Figure~\ref{VIRcomp}:
%note in particular the 
holes in the parent catalog and the varying
sampling density in the spectroscopic data, due to the multiple
passes over the central area.  The ``striping'' effect  due to
%of 
the
slit-placing software is not obvious
%evident 
at this resolution and is better
appreciated in Figure~\ref{SSfield}, where only one quadrant is displayed.

%\section{Simulated VVDS Surveys}
\section{Constructing mock VVDS surveys}

The only way to understand the relative importance of the
selection biases discussed above and test possible correction schemes
is to create and analyze realistic simulations of our survey.  
Provided these simulations are realistic enough, they allow us (1) to understand quantitatively the magnitude of observational biases on the final statistical quantities to be measured, and (2) to estimate realistic errors that include cosmic variance. Both these points require that mock observations contain a spatial distribution of galaxies consistent with VVDS observations -- so as to measure clustering and cosmic variance -- along with realistic photometric and physical properties of simulated galaxies -- so as to mimic selection effects. The \galics{} model for galaxy formation (\cite{gal1}) along with the \momaf{} mock observing tool (\cite{gal2}) fulfill these requirements and we thus use them to build ``pre-observation'' catalogs that we then ``observe'' by progressively adding all the VVDS observational biases and selections.

In this section, we first describe the \galics{} simulation that we use, before discussing how we build simulated VVDS observations that account for all identified biases.

% In this
%section we shall describe the construction of
%Our goal is to construct 
%a set of mock $1\times 1$ square-degrees, realistic Deep VVDS surveys
%based on the so-called \galics simulations (\cite{gal2}).  The
%advantage of the mock samples is twofold: 1) understand quantitatively
%the magnitude of observational biases on the final statistical
%quantities to be measured; 2) }
%with all the above effects accurately reproduced, studied and
%corrected.  Having a sufficient number of independent realizations of
%the survey will, in addition, allow us to 
%estimate realistic errors
%that will include -- assuming that the power spectrum of the
%simulations is a good description of the cosmic one -- the contribution
%from cosmic variance, which 
%is normally neglected 
%cannot be accounted for by ``internal'' techniques like, e.g., the
%bootstrapping. 
%{\bf On the other hand, it is worth to note that to 
%reach both these goals we do not need to have a perfectly realistic
%simulation. To understand the observational biases and to test the
%correcting methods we 

\subsection{The \galics{} simulations}

\galics{} (for {\it Galaxies In Cosmological Simulations}, see \cite{gal1}) is a model of hierarchical galaxy formation which combines high resolution
cosmological simulations to describe the dark matter content of the
Universe with semi-analytic prescriptions to deal with the baryonic
matter.

The cosmological N-body simulation we refer to throughout this paper
assumes a flat cold dark matter model with a cosmological constant
($\Omega_m = 0.333$, $\Omega_{\Lambda} = 0.667$). The simulated volume
is a cube of side $L_{box}=100h^{-1}$Mpc, with $h = 0.667$,
containing $256^3$ particles of mass $8.272\times 10^9$M$_{\odot}$,
with a smoothing length of 29.29 kpc. The power spectrum was set in
agreement with the present-day abundance of rich clusters ($\sigma_8 =
0.88$, from \cite{Eke96}), and the DM density field was evolved from z=35.59 to z=0,
outputting 100 snapshots spaced logarithmically in the expansion factor.

\galics{} builds galaxies from this simulation in two steps. First, halos of DM containing more than 20 particles are identified in each snapshot using a friend-of-friend algorithm. Their merging history trees are then computed following the constituent particles from one output to the next. Second, baryons are evolved within these halo merging history trees according to a set of semi-analytic prescriptions that aim to account for e.g. heating and cooling of the gas within halos, star formation and its feedback on the environment, stellar population evolution and metal enrichment, formation of spheroids through galaxy mergers or disc instabilities.

Three main points make \galics{} particularly suitable for this study. First, this model yields a wide range of predictions, among which luminosities (in many bands from the UV to the sub-mm), physical properties (such as sizes of galaxies), and the positions of galaxies within the simulation snapshots. Second, these properties have been shown to be in a rather good agreement with various observations (e.g. \cite{gal1}, \cite{gal3}). Third, mock observations are readily available from the GalICS Project's web-page\footnote{{\tt http://galics.iap.fr}}. These mock observations include 50 catalogs of $1\times 1$ sq. deg. that contain all the information we need in this study: apparent magnitudes in the BVRI filters used at the CFHT, apparent sizes of the galaxies, angular coordinates in the mock sky, and redshifts.

\begin{figure} \centering
   \includegraphics[width=8.5cm]{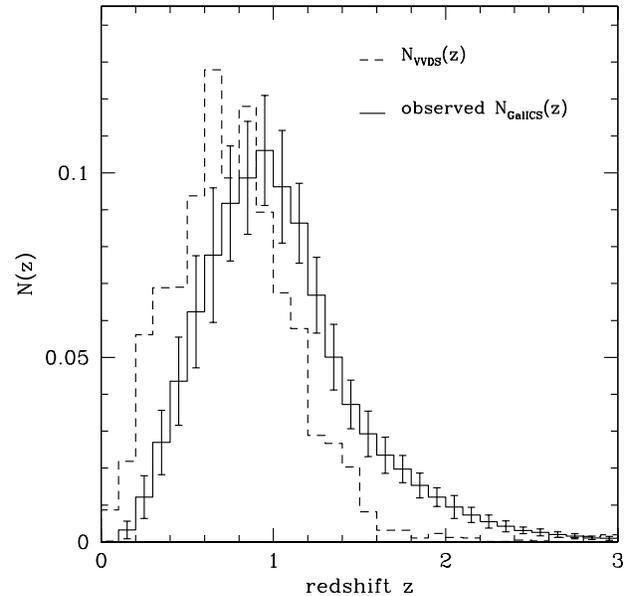} \caption{
   Average redshift distribution in the $50$ mock VVDS-02h surveys,
   normalized by the number of objects in each cone, compared to the
   redshift distribution of the observed VVDS galaxies. Note how
   the semi-analytic model of galaxy formation used to construct the
   \galics{} simulations differs from the real data.  This is not a
   concern for the purposes of this work:  first, we are performing
   {\it internal} tests of the effect of observing biases and on their
   correction, which depends on the small-medium scale clustering
   properties.  Second, when error bars are estimated for a specific
   redshift slice, their amplitude is re-normalized accordingly, to
   account for the different number of galaxies. } \label{nofz}
   \end{figure}

Before using \galics{} mock samples, it is useful to state their limitations (see however \cite{gal2}, for a thorough description of these). There are mainly three shortcomings to mock catalogs made using \galics{}. First, because of the finite mass resolution of the root simulation, faint galaxies are not well described, or even missed when they lie in unresolved haloes. This is not an issue for the present study, however, because the VVDS detection limit is brighter than \galics{}'s resolution. Second, because mock catalogs are built from a simulation of a finite volume, they involve replications of this volume, along and perpendicular to the line of sight. These replications lead to some negative bias in the correlation functions, of at most $\sim 10$\%. This is not a concern in this paper, because we just need an approximate match with the observed data in order to perform an \emph{internal} consistency check. \galics{} catalogs do provide an adequate match. Third, the mock catalogs do not describe density fluctuations on scales larger than the size of the simulated volume ($\sim 100$ h$^{-1}$Mpc). This implies that cosmic variance estimates are likely to be under-estimated if the volume probed by a mock catalog is larger than the simulated volume. This under-estimate, however, depends on the galaxy population considered: it will be large for rare objects and small for ``normal'' galaxies. In other words, because cosmic variance is basically given by the integral of the correlation function over the survey, the error on the estimated cosmic variance depends on how much of this integral we miss, that is, on how strongly the studied galaxies are clustered. From Fig. 9, it can be seen that the size of the simulation is enough for this under-estimate to be small at the scales we consider (i.e. from $0.1$ to $10$ h$^{-1}$Mpc). The dispersion found among the 50 \galics{} cones is thus expected to be a good estimate of cosmic variance.
The mean number of galaxies with $17.5 < I_{AB} < 24$ in the artificial catalogs is $77396$. The average redshift
distribution of these $50$ cones is shown in Figure~\ref{nofz}, along with
the VVDS first epoch $N(z)$ (\cite{LeFevre2004}).

%Using the \galics simulation, we have been able to build $50$
%quasi-independent mock surveys, each corresponding to a sky area of
%one square degree and containing a mean of $77396$ galaxies in each,
%with $17.5 < I_{AB} < 24$. {\bf HERE TOO NEEDS TO BE ADDRESSED CLEARLY
%HOW INDEPENDENT THESE CONES ARE.  WE MUST BE QUANTITATIVE HERE.
%JEREMY IS THE ONE TO PROVIDE THIS INFORMATION} The average redshift
%distribution of these $50$ cones is shown in Figure~\ref{nofz}, where
%we also make a comparison to the VVDS first epoch $N(z)$
%(\cite{LeFevre2004}. 
%
We note that the redshift distribution of the simulated
galaxies differs significantly from that observed by the VVDS for the
real Universe.  This is simply telling us that the semi-analytic
galaxy formation model adopted to construct the \galics{} simulations,
while adequately reproducing a number of observed features (see
\cite{gal2}) is not 100\% correct.  This, however, is of no importance for
the current analysis, as our main goal is to test the internal differences in
the measured quantities when either the original parent sample or the
final spectroscopic sample are observed.  The accuracy of these tests
depends essentially on the small-scale properties of the simulated
galaxies (like the mean inter-galaxy separation and clustering),
rather than on the global redshift distribution.  Conversely, in the
estimate of error bars the difference in absolute numbers between the
real and simulated samples within a given redshift slice will clearly 
have to be taken into account.  

%These mock surveys can now be
%``observed'', applying the selection function that we have described
%in the previous section.

\subsection{CCD photometric mask}

Bright (often saturated) stars represent a practical obstacle to
accurate galaxy photometry and their diffused light
can affect large areas of a CCD astronomical image. All
such areas were excised from the VVDS photometric catalogs: there
are no sources
in these regions (\cite{hjmcc}). Similarly, a
``dead'' area in the 02h field has been produced by a beam of scattered
light that crosses a large part of the field from North-East to
South-West. In total, a few percent of the total area are lost due to
these defaults. The information on these ``holes'' in the photometric
catalog is stored in a FITS binary mask, with null values corresponding
to dead pixels.  We have used this mask on the 
mock samples to exactly reproduce the pattern of
the observed data in our simulations.

\subsection{Effect of galaxy angular sizes}

In order to maximize the number of spectroscopic targets, the Super-SPOC
software (\cite{SSPOC}) makes a choice of a targeted galaxy based also on the
galaxy projected angular radius along the slit direction.  This means
that smaller galaxies are sometimes preferred as they allow the program
to increase the number of targets.  Any realistically simulated
spectroscopic sample must take this into account. Therefore, we have
computed for each simulated galaxy in \galics{} a realistic angular
radius, using the following procedure.

\galics{} describes galaxies with three components~: a disc, a bulge and possibly a nuclear starburst. For each of these, the model predicts the mass and a scale-length that assumes the disc is exponential while the other two spheroidal components follow a Hernquist profile (\cite{Hernquist}). We used these sizes to define an overall radius for each galaxy, which encloses 90\% of the total mass. Assuming that light has the same distribution as mass, we then convert this radius to an apparent angular size, assuming the above-mentioned cosmology.

\subsection{Artificial stars}

The VVDS spectroscopic targets are selected purely on magnitude,
$I_{AB}\le 24$ and $I_{AB}\le 22.5$ in the Deep and Wide parts of the
survey, respectively, without any {\it a priori} star-galaxy
separation. This avoids biases against compact galaxies
and AGNs which may be introduced at faint
magnitudes by unreliable star-galaxy classification
based on morphology. Consequently, our
spectroscopic sample is contaminated by stars. About
$8.5\%$ of the collected spectra in the VVDS-Deep are stars 
and are discarded (%although 
the exact
number %depends 
depending on galactic latitude 
%and 
can be as high as 20\% in some
cases for the ``Wide'' survey).  
These stars obviously have no impact on the clustering
analysis.  Their only effect is to reduce the total number of
targeted galaxies, thus slightly affecting the overall statistics by 
increasing the expected variance.
Since our aim here is to precisely quantify the biases and
uncertainties on galaxy correlations computed from the final
spectroscopic sample, and compare them to the original parent sample,
we decided to also take into account this small contribution. 
We therefore added to the artificial survey fields a set of
simulated stars.

\begin{figure} \centering \includegraphics[width=8.5cm]{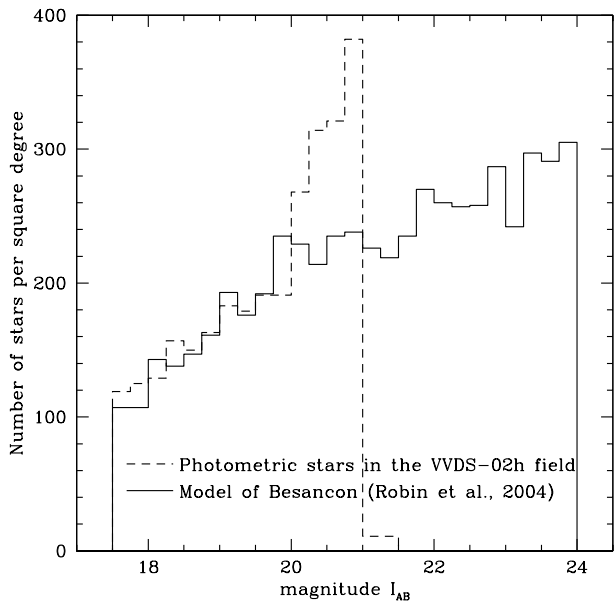}
   \caption{Number counts of artificial stars added to the \galics{}
   simulation, compared to the actual counts of stars in the VVDS-02h
   field, identified morphologically from the photometric data. The
   excess in the VVDS above $I_{AB}=20$ is 
%readily explained by
due to the inability of the morphological compactness criteria to
   discriminate stars from galaxies and QSOs at faint magnitudes. 
   When this is taken into account,
%At bright magnitudes, 
the models from \cite{Robin} reproduce very well the actual distribution of
stellar objects in the VVDS.}  
\label{stars}
\end{figure}

Using the on-line tool of \cite{Robin} \footnote{The \textit{Model of
    stellar population synthesis of the Galaxy} developed by
  \cite{Robin} produces a reliable catalog of stars with appropriate
  number counts and magnitudes in the visible and near-infrared
  spectral ranges in the Johnson-Cousins and Koornneef systems,
  respectively.}  we created a one-square-degree catalog of artificial
stars with $17.5$ $I_{AB}$$\le 24$, which was added to the artificial galaxy
photometric catalogs.  Figure~\ref{stars} shows the number counts of
the added stars, compared to the observed distribution at bright magnitudes
in the 02h field (as identified by {\sl S-extractor},
\cite{sextractor}).  The observed excess above $I_{AB}=20$ in the 02h
field is the effect of mis-classified galaxies and QSOs, which also
corroborates our choice of excluding any pre-selection for the VVDS
spectroscopy, to avoid throwing these objects away.

As this parameter is used by VMMPS, apparent angular
radii have also been assigned to artificial stars, using the observed
distribution of stellar sizes in the 02h field, identified
photometrically down to $I_{AB}=21$ and spectroscopically at fainter
magnitudes. This range of apparent stellar radii corresponds to the
sizes of the point spread function (``seeing'') at the faint Kron
radii measured for stars by {\sl S-extractor}.

\subsection{Spectroscopic success rate}

\begin{figure} \centering
\includegraphics[width=8.5cm]{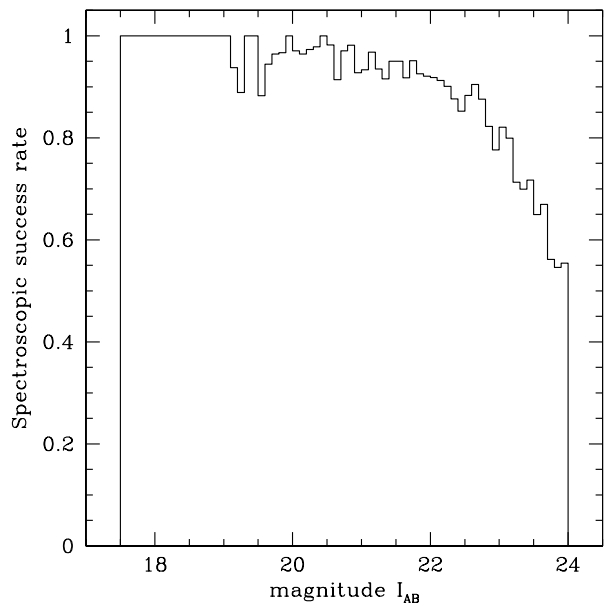}
\caption{
%Measured 
Spectroscopic success rate  per magnitude bin in the VVDS 02h field,
including only those redshifts used for the clustering analysis 
%{\it R! with a confidence level greater than $80\%$.} R!
}
\label{success} 
\end{figure}

Objects selected by the slit-positioning code do not yet form the final
redshift catalog.  For some of the objects, redshift measurements are
impossible, usually because of poor signal-to-noise. This
incompleteness is clearly a function of magnitude.  We define the
spectroscopic success rate as the ratio of the number of redshifts
used for clustering analysis
%
%{\it R! measured with sufficiently high confidence (for the clustering
%analysis, we use redshifts with flag $\geq2$, i.e. with a $\ge 80\%$
%probability to be correct,  see \cite{LeFevre2004}) R!} 
%
to the total
number of spectroscopically observed objects. Figure~\ref{success}
shows the spectroscopic success rate as a function of magnitude, which
corresponds in practice to the probability of measuring the correct
redshift of a galaxy as a function of its magnitude in the current
observational configuration. Overall, this shows that we are able
to obtain a redshift for more than $80\%$ of the targeted objects 
%down to
between $I_{AB}=17.5$ and $24$. 
We therefore apply this same probability function to
our mock ``observed'' catalogs, rejecting the corresponding 
fraction of targeted objects.  We make the simplifying assumption
that the spectroscopic success rate is the same for all galaxy types.

\subsection{VIMOS spectral resolution}

The last point to be taken into account to produce a fully realistic
mock redshift catalog is the resolution of the VIMOS spectrograph in
the set-up used for the VVDS (Low-resolution RED Grism, $R\simeq 230$)
which translates into a typical {\it rms} error on the measured
redshift which is around $\sigma_{cz}\simeq 275$ km/s.  We
therefore added to the final set of mock redshifts a
Gaussian-distributed dispersion with the same {\it rms} and zero mean.

\subsection{Overall properties of mock VVDS surveys}

All of the steps described above have been applied to each of the $50$
one-square-degree \galics{} surveys, producing a corresponding number of
mock redshift samples which reproduce with fidelity the lay-out,
properties and biases of the first-epoch VVDS 02h sample.

\begin{figure}
   \centering
   \includegraphics[width=8.5cm]{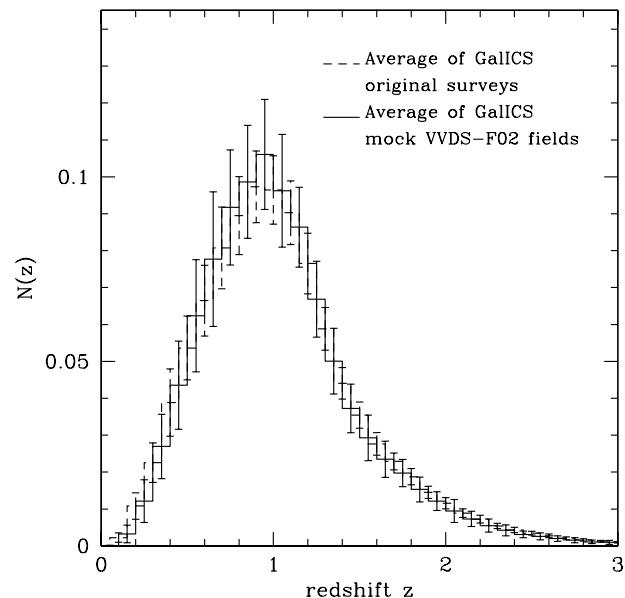}
   \caption{Average redshift distribution in the \galics{} mock catalogs before and after
     the full observing strategy is applied.  No bias in the redshift
     distribution is observed. }
        \label{Nzcomp}
   \end{figure}

Figure~\ref{Nzcomp} shows that, despite the slight bias of SSPOC
towards choosing smaller (and therefore fainter) objects, the redshift
distribution $N(z)$ of the final spectroscopic samples is unbiased with
respect to the original complete \galics{} one-square-degree survey.
The difference observed in Figure~\ref{nofz} between the
original and observed
%\galics{} 
simulated cones 
%and the observations 
is %mainly 
therefore only the result
of the model of galaxy formation adopted for the simulation,
and not %to 
of a selection effect.  
%In other words, Figure~\ref{Nzcomp}
%suggests that the observed $N(z)$ of the current 02h field, should also
%be an unbiased sample of the true underlying redshift distribution of
%the complete sample, at least to $z\sim 2$.  Since we introduced into 
%the mock samples a success rate being a function of apparent magnitude 
%(for fairly sure redshifts), as described by the curve of Figure~\ref{success}, 
There was no way we could introduce, e.g., a stronger incompleteness 
in the final $N(z)$ at $z>1$.

%GGG: removed as repetition of what said in 3.1
%
%We note that the intrinsic difference in the $N(z)$ of the data and
%the simulations does not influence our internal consistency tests on
%correlation function estimates, but will have to be taken into account
%when computing error bars from the mock catalogs within redshift bins
%with significantly different statistics. 

%\section{Biasing Effects on Two-Point Correlation Statistics}
\section{Two-point correlation statistics}

\subsection{General estimator}
The two-point correlation function \xir is defined as the excess
probability above random that a pair of galaxies is observed at a given
spatial separation $r$ (\cite{peebles80}).

It is the simplest statistical measurement of clustering, as a function
of scale, and it corresponds to the second moment of the distribution.
Various recipes have been proposed to estimate two-point correlation
functions from galaxy surveys, in particular to minimize the biases
introduced by the finite sample volume, edge effects, and photometric
masks (\cite{Hamilton93}, \cite{lansal}).  
Here we adopt the Landy-Szalay estimator, 
that expresses \xir as

\begin{equation}
\xi(r) = \frac{N_R(N_R-1)}{N_G(N_G-1)} \frac{GG(r)}{RR(r)} 
        - 2 \frac{N_R-1}{N_G} \frac{GR(r)}{RR(r)} + 1 \,\,\,\,\,   .
\label{lseq}
\end{equation}
In this expression, $N_G$ and $N_R$ are the mean density (or, equivalently, the total
number) of objects respectively in the galaxy sample
%; $N_R$ is the mean density of 
and in a catalog of random points distributed within the same survey
volume and with the same redshift distribution and angular selection biases;
$GG(r)$ is the number of independent galaxy-galaxy pairs with
separation between $r$ and $r+dr$; $RR(r)$ is the number of independent
random-random pairs within the same interval of separations and $GR(r)$
represents the number of galaxy-random pairs.

%{\bf In the following section we shall apply the formulae and estimators described in this section
%to the whole set of  

%{\bf FORMER 4.3.1, 4.3.2 AND 5.1 MOVED HERE}

\subsection{Redshift-space correlations}
We know that the three-dimensional galaxy distribution recovered from a
redshift survey is distorted due to the effect of peculiar velocities.
For this reason, the redshift-space separation $s$ differs from the
true physical comoving separation $r$ between two galaxies.  Since
random velocities affect only redshift and not position on the sky, the
stretching occurs only radially. Redshift distortions can be measured
and separated from true spatial correlations by computing the function
$\xi(r_p,\pi)$, where the separation vector of a pair of galaxies $s$
is split into two components: $\pi$ and $r_p$, respectively parallel
and perpendicular to the line of sight. Given two objects at redshifts
$z_1$ and $z_2$, with observed radial velocities $v_1=c z_1$ and $v_2=c
z_2$ ($c$ being the speed of light), we can define (\cite{Fisher94})
the line of sight vector $\vec{l} \equiv (\vec{v1}+\vec{v2})/2$ and the
redshift difference vector $\vec{s} = \vec{v1} - \vec{v2}$, and also:

\begin{equation}
\pi \equiv \frac{\vec{s} \cdot \vec{l}}{H_0 |\vec{l}|}, \hspace{1cm}
r_p^2 \equiv \frac{\vec{s} \cdot \vec{s}}{H_0^2} - \pi^2. 
\end{equation}
Equation~\ref{lseq} can be generalized to the case of $\xi(r_p,\pi)$,
if we count the number of pairs in a grid of bins $\Delta r_p$ 
and $\Delta \pi$ instead of singular bins $\Delta r$ or $\Delta s$.

%In $\xi(r_p,\pi)$ we can measure different kinds of redshift distortions.
Observed distortions in galaxy surveys can be separated into two main
contributions: on small scales, the distortion is dominated by random
internal velocities in groups and clusters, causing a stretching of
$\xi(r_p,\pi)$ along the $\pi$ direction (the so-called ``fingers of
God'' effect). On large scales, on the other hand, $\xi(r_p,\pi)$
contours tend to be flatter, due to coherent infall of galaxies onto
large-scale overdensities, known as the ``Kaiser effect''
(\cite{kaiser}).  The latter is a weak effect and needs very large
samples to be seen with sufficient accuracy, as shown by the 2dF survey
(\cite{2DFGRS}).

\subsection{Projected correlation function $w_p(r_p)$}

We can recover the real-space correlation function $\xi(r)$ by
projecting $\xi(r_p,\pi)$ along the line of sight, onto the $r_p$
axis.  In this way we integrate out the dilution produced by the
redshift-space distortion field and obtain a quantity, $w_p(r_p)$,
which is independent %of
of the redshift-space distortions:

\begin{equation}
w_p(r_p) \equiv 2 \int_0^\infty \xi(r_p,\pi) dy = 2 \int_0^\infty  
\xi_\circ\left[(r_p^2+y^2)^{1/2}\right] dy.
\label{wpdef}
\end{equation}
In the right-hand side of the equation, $\xi_\circ$ is simply the usual
real-space two-point correlation function $\xi(r)$, evaluated at the
specific separation $r=\sqrt{r_p^2+y^2}$.  If we now assume a
power-law model
\begin{equation}
\xi(r) = \left(\frac{r}{r_0}\right)^{-\gamma},
\label{powlaw}
\end{equation}
with $\gamma$ being the slope of the correlation function and 
$r_0$ the correlation length,
the integral can be computed analytically, giving as a result

\begin{equation}
w_p(r_p) = r_p \left(\frac{r_0}{r_p}\right)^\gamma 
\frac{\Gamma\left(\frac{1}{2}\right)\Gamma\left(\frac{\gamma-1}{2}\right)}
{\Gamma\left(\frac{\gamma}{2}\right)},
\label{wpmodel}
\end{equation}
where $\Gamma$ is Euler's Gamma function. 

%GGG
\section{Error estimate and fitting technique}
\label{sec_error}
%GGG
%The first part of this paper has been dedicated to the correction of
%instrumental and observational biases, affecting the correlation
%function measurements.   
%The 
%second 
%{\bf first classical use} use of 
%the VVDS 
%mock surveys is that of estimating realistic error bars for those
%statistical quantities that are being measured from the
%VVDS,
%{\bf actual survey} under {\bf hopefully identical} 
%%similar 
%conditions.  

\subsection{Error bars on correlation functions}

Ideally, if the studied data set consisted of a large enough number
of statistically independent pairs, such that the central limit theorem
applies, then the distribution of estimates of $\xi$ in an ensemble of
similar samples should be Gaussian. The $1 \sigma$ uncertainty --- the
``cosmic error''--- in $\xi$ would then be the square root of its
variance $< \Delta \xi^2 >$ (\cite{peebles73}). However, the
theoretical expression for $< \Delta \xi^2 >$ depends on the poorly
known and difficult to measure four-point correlation function.
Moreover, since the measured $\xi$ is not exactly coincident with the
theoretical $\xi$, we expect its uncertainty to be
% the measured $\xi$ may have its uncertainty 
also somewhat different from the value provided by the
theory. This effect is known as a cosmic bias.

A few different ways of
estimating errors on two-point correlation functions have been used in
the literature (for a wider discussion, see e.g. \cite{Hamilton93},
\cite{Fisher94}, \cite{bcgs01}).  
The case closest to the ideal situation is when the survey is large
enough that it can be split into a number of sub-samples.
Correlations are then estimated independently for each of these, and
error bars for the parent sample computed as the {\it rms} values.
This has been for example the case of the angular correlation function
from the APM survey (e.g. \cite{maddox}).  However, the number of
sub-samples cannot be large, otherwise the explored scales will be
significantly reduced with respect to the parent survey.  The
consequence is that the variance is typically overestimated and these
represent usually upper limits to the true errors.

Simple Poissonian errors (e.g. proportional to the square root of the
total number of galaxy pairs in each bin) underestimate the error bars
substantially.  Statistical corrections were proposed
(\cite{kaiser86}) by multiplying Poissonian errors by a factor $1+4
\pi n J_3$, with $n$ being the number density of objects and $J_3 =
\int^{r_j} r^2 \xi(r) dr$, where we assume that the actual correlation
function vanishes for $r \geq r_j$. However, this method also tends to
give relatively small errors (\cite{Fisher94}).

%I think bootstrap resampling has been largely discredited as 
%means of estimating errors in the galaxy correlation functions, in
%favour of either analytic prescriptions like those from mr. Colombi,
%or alternatively, as you say, large n-body simulations. I would cut
%this paragraph down to the absolute minimum!  I
%would

Over the last twenty years a widely used method has been the
so-called ``bootstrap resampling'' 
(\cite{barrow84}).  It is based on the idea of ``perturbing'' the data
set, by randomly creating a large number of comparable ``pseudo
data-sets'', which differ only slightly from the original sample.
If this contains $N$ objects, then each bootstrap sample is created
selecting $N$ of these, but allowing for multiple selections of the
same object.  This means that some objects will not be included in one
given pseudo data-set, while others will be counted twice or three times.
This is a good test of the robustness of measured correlations,
especially on large scales where having a large number of pairs does
not always mean a robust measurement: consider for example the case of 
a single isolated galaxy at a separation of $\bar r$ from a cluster 
containing 1000 galaxies.  $\xi(\bar r)$ will contain a large number
of pairs, however only one will be independent.  On the other hand,
bootstrap errors often tend to over-estimate the theoretical variance
$< \Delta \xi^2>$.  In general, however, despite debates on their
theoretical justification, they have represented a practical way to
obtain error bars in correlation analysis which are not far from the
true ones.  

%I am not sure if I agreed with that -- the boostrap errors seem to be
%much too large. Anyway, they are of course almost useless because the
%neglect the cosmic errors. 
The use of bootstraping became less and less popular in recent years,
with the advent of large N-body simulations, reproducing the matter
distribution over significant volumes of the Universe.  Coupled to
physically sound definitions of ``galaxies'', these allowed the
construction of sets of independent mock surveys, from which ensemble
errors could be computed from the scatter in the different catalogs.
This is the same technique used to
construct our VVDS mock surveys. Clearly, a good match is necessary
between the volume and resolution of the simulation, on one side, and
the depth and size of the survey on the other.  Furthermore, the power
spectrum of the simulation must provide a realistic description of long
waves, so to properly include cosmic variance.  Progress both in our
knowledge of structure on the largest scales and in the size and
resolution of N-body simulations has improved on
early applications of this technique (\cite{Fisher94}).   For this
reason, since the \galics{} simulations are available, we could use
this as our main method for error estimation.

However, as we detail below, the
covariance matrix reconstructed from the simulations cannot be
applied in a straightforward way to the observed data.  Indeed, our
fitting technique, discussed below, handles the covariance
matrix to properly account for bin-to-bin correlations when fitting
correlation functions: when the covariance matrix extracted from the
set of 50 mock VVDS surveys is used (after proper normalization of the
average values), the fit is often unstable.
%evidently wrong.  
In other words,
the covariance matrix produced by the ensemble of mock surveys,
although providing sufficiently realistic diagonal elements, has
off-diagonal non-zero values which differ from those pertaining to the
data sample (which of course are unknown).  For this 
reason, we modified our strategy and resort to the bootstrap
technique {\it to estimate the bin-to-bin
covariance}.  This means that our error bars on the estimated
correlation functions are obtained via the more reliable
scatter between the mock surveys, but a bootstrap is used to estimate
the off-diagonal terms of the covariance matrix.

\subsection{Fitting correlation functions}
\label{fit}
%GGG
It is well known that fitting of correlation functions like \xis
or $w_p(r_p)$  cannot be
performed via the standard least-squared method, due to the
correlation existing among the different bins.  
The procedure we adopted to estimate the power-law parameters of
$\xi(r)$, $r_0$ and $\gamma$ from the projected function
$w_p(r_p)$, using eq.~\ref{wpmodel} 
%GGG
follows Fisher et al. (1994) and Guzzo et al. (1997), with some
specific differences that are described in the following.

Let us consider a given redshift slice $[z_1 - z_2]$.  Within this
same interval, we estimate the correlation function
$\xi(r_p,\pi)$ from: 1) $50$ mock VVDS surveys; 2) the real VVDS data;
3) $N_{boot}$ (typically $100$) bootstrap resamplings of the VVDS data.
We then compute, for each of these estimates, $w_p(r_p)$, projecting
$\xi(r_p,\pi)$ along the line of sight (eq.~\ref{wpdef}), with an
upper integration limit $\pi_{max}$, chosen in practice so that it is
large enough to produce a stable estimate of 
$w_p$. Similarly to other authors (see e.g. \cite{G97}), we find
$w_p(r_p)$ quite insensitive to the choice of $\pi_{max}$ in the range
of $15~{\rm h}^{-1} {\rm Mpc} < \pi_{max} < 25~{\rm h}^{-1} {\rm Mpc}$
for $r_p < 10~{\rm h}^{-1} {\rm Mpc}$.  
Too small a value for this limit would miss small-scale power, while 
too large a value has the effect of  adding noise into $w_p$. After a
set of experiments we have chosen $\pi_{max}=20~{\rm h}^{-1} {\rm
Mpc}$.

In the following, we call $w_p^k(r_i)$ the value of $w_p$, computed
at $r_p=r_i$ in the cone $k$, where $1 \le k \le N_{\galics}=50$ if we
consider the \galics{} data or $1 \le k \le N_{boot}$ if we
consider the bootstrap data.  
If not otherwise mentioned, $N_{boot}=100$ is used.

Whether we 
consider the mock or bootstrap samples, 
we can always compute the associated covariance matrix, ${\bf C}$,  between the
values of $w_p$ in $i^{th}$ and $k^{th}$ bins:
\begin{eqnarray}
 {\bf C}_{ik} & = & \langle\left(w_p^j(r_i) - 
\langle w_p^j(r_i)\rangle_j \right) \left( w_p^j(r_k) - 
\langle w_p^j(r_k)\rangle_j\right)\rangle_j, 
\end{eqnarray}
where '$\langle \rangle_j$' indicates an average over all bootstrap or
mock realizations.  
When the correlation function is computed from a finite sample, the
values of $\xi(r)$ (or $w_p(r)$) at different separations are not
independent\footnote{For example, imagine that one galaxy is removed 
from the sample: this galaxy contributes pairs at many different
separations, thus affecting virtually all bins in the
correlation function.}
%, being a part of pairs at many different
%separations; therefore, all the bins will be affected by its removal.}.  
For this reason
one cannot use a straightforward $\chi^2$ minimization --- which
assumes that all bins are independent and that the errors follow 
the Gaussian distribution ---
to find the best-fit parameters of a model to the
observed data.   However, {\bf C}  is symmetric and real and therefore can be
diagonalized by a unitary transformation if its determinant is
non-vanishing.  In practice, the estimated functions are oversampled,
{\bf C} is not singular and therefore can be inverted by a simple
Cholesky decomposition (\cite{NumRecip}, Volume 1, Chapter 2)\footnote{Note that if the number of
bins we want to fit, i.e. the size of the matrix, were greater or
equal to the number of realizations then, even if the matrix remains
symmetric, the vectors would not be independent and the matrix {\bf C}
could not be inverted.}.  Then,
%GGG: moved up   (see also \cite{Fisher94} and \cite{G97} for a
%similar approach), 
if we now call ${\bf H}
={\bf C}^{-1}$,
we can fit $w_p^{VVDS}$ by minimizing a generalized $\chi^2$, which is
defined as 
\begin{eqnarray}
\chi^2 &=& \sum_{i=1}^{N_D}\sum_{j=1}^{N_D} \left(w_p^{mod}(r_i)
\right. \\ 
& & - \left.
w_p^{VVDS}(r_i)\right)H_{ij}\left(w_p^{mod}(r_j)-w_p^{VVDS}(r_j)\right),
\nonumber
\end{eqnarray}
as a function of the two free parameters $r_0$ and $\gamma$ of
$w_p^{mod}(r_p)$.

In principle, the complete process could be done using only our set of
50 mock VVDS surveys.   In practice, as explained above, 
%GGG to avoid problems with proper estimating of 
the bin-to-bin covariance obtained from the \galics{} mock samples
does not provide a statistically stable matrix to be used
with the generalized $\chi^2 $ method. Therefore, we most appropriately used
% it is better to use 
the covariance 
matrix obtained from the $N_{boot}$ bootstrap resamplings
of the galaxy data set.  
% At the same time, {\bf however}, we use 50 mock
%surveys to obtain the most realistic error contours on our estimated
%$(r_0,\gamma)_{data}$, as these --- unlike bootstrap errors --- include
%cosmic variance.  The final error contours, therefore, are
%obtained after re-normalization with respect to the different central 
%$(r_0,\gamma)$ values found in the data and in the simulation.
%{\bf BM: A SENTENCE OF A COMMENT HOW THE RE-NORMALIZATION IS DONE}

This provides the best solution for $(r_0,\gamma)_{data}$ that minimizes 
the error contour
$\chi_{boot}^2(r_p,\gamma)$. At the same time, however, we use $50$ mock
surveys to obtain the most realistic error contours $\chi^2(r_p,\gamma)$ on our estimated
$(r_0,\gamma)_{data}$, as these - unlike bootstrap errors - include cosmic
variance.

The final error contours, therefore, are obtained fitting the mean of the
$50$ $w_p$ mock VVDS surveys, using a covariance matrix computed from the same
$50$ $w_p$.
This process provides a solution for $(r_0,\gamma)_{GalICS}$ associated with the
error contours $\chi_{GalICS}^2(r_p,\gamma)$. We then re-center these contours
around $(r_0,\gamma)_{data}$ with the renormalization
$r_p \leftarrow r_p \times (r_0^{GalICS}/r_0^{data})$ and
$\gamma \leftarrow \gamma \times (\gamma^{GalICS}/\gamma^{data})$.

To take into account the different $N(z)$ of GalICS and VVDS, we multiply
the error contour $\chi_{GalICS}^2$ computed for each redshift slice by 
a factor $N_{VVDS}/N_{GalICS}$, where $N_{VVDS}$ is the number of VVDS galaxies
and $N_{GalICS}$ is the number of \galics{} galaxies in this redshift slice.

The error bars  computed as above for each $w_p(r_i)$ value correspond
to the {\it rms} of the $50$ $w_p^k(r_i)$, normalized to the data. 

\section{Biasing effects and their removal}

We now quantitatively
establish the impact of the VVDS selection effects on the measured
correlations and the accuracy of our correcting scheme, using the
\galics{} mock samples.

%GGG
\subsection{Impact on angular correlations}

   \begin{figure} \centering
   \includegraphics[width=8.5cm]{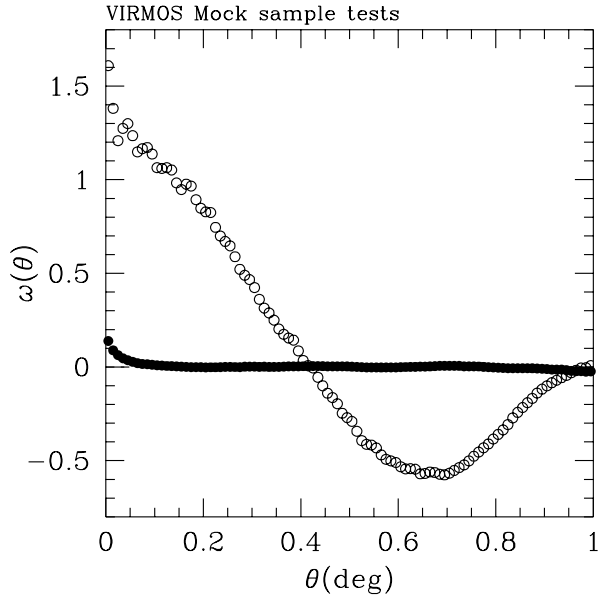} \caption{Impact of
   the observational process on the estimate of the angular two-point
   correlation function $\omega(\theta)$ for one mock VVDS survey
   (open circles), compared to that of the original parent field
   (filled circles), for one mock VVDS cone. The large
   distortion, introduced by the observing strategy 
%into the two-point correlation function, is visible 
affects practically all angular scales.}  \label{ang1} \end{figure}
    
As we have seen in the previous section, the biases and selection
effects due to the observing strategy and instrumental limitations
%mostly 
affect the properties of the angular distribution of
objects, with respect to a random sub-sampling of 
%the complete galaxy distribution
galaxy clustering process.  It is therefore the angular correlation
function $\omega(\theta)$ that will primarily reflect these biases.
Clearly, there is no specific scientific reason to measure the
angular correlation function from the spectroscopic sample, as this
can be done more easily and with much greater confidence using the
full VVDS photometric catalog (\cite{hjmcc}). $\omega(\theta)$
allows us to illustrate the level
% Our goal here is to investigate the kind 
of distortions introduced by our angular
selection function.  
%It is also clear that the ability to recover
%sufficiently well the original shape and amplitude of
%$\omega(\theta)$ will be a strong test of any correction scheme to
%be used for recovering the spatial two-point function, which is our
%main goal here (see \S~\ref{correction}).

To this end,
figure~\ref{ang1} shows the angular correlation function computed
% , for illustrative purposes, what happens when we compute $\omega(\theta)$ 
from one mock VVDS redshift survey without
correcting for these effects (i.e. using a random sample which simply
follows the geometrical borders of the galaxy sample, as one would do
for a homogeneous angular selection), compared to that of the original
mock catalog.  We used 
%$\omega(\theta)$ is computed here using 
the angular version of the
%the same
Landy-Szalay estimator (eq.~\ref{lseq}), 
%in its angular version,
without taking into account any incompleteness on any scales.  The
comparison to the parent survey $\omega(\theta)$ reveals the very
strong distortions introduced over a wide range of angular scales.

%______________________________________________
   \begin{figure*}
   \centering
   \includegraphics[width=8.5cm]{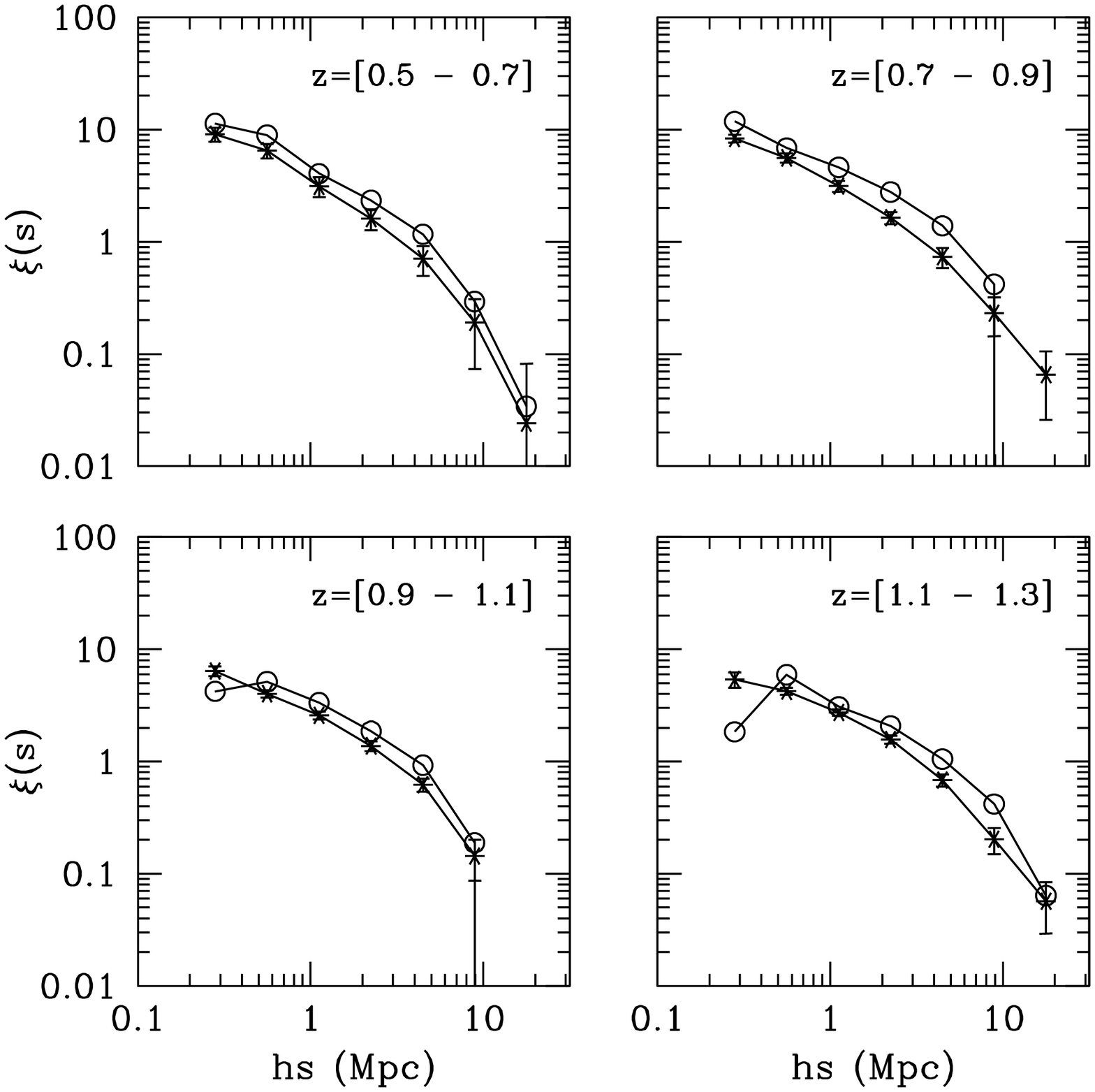}
   \includegraphics[width=8.5cm]{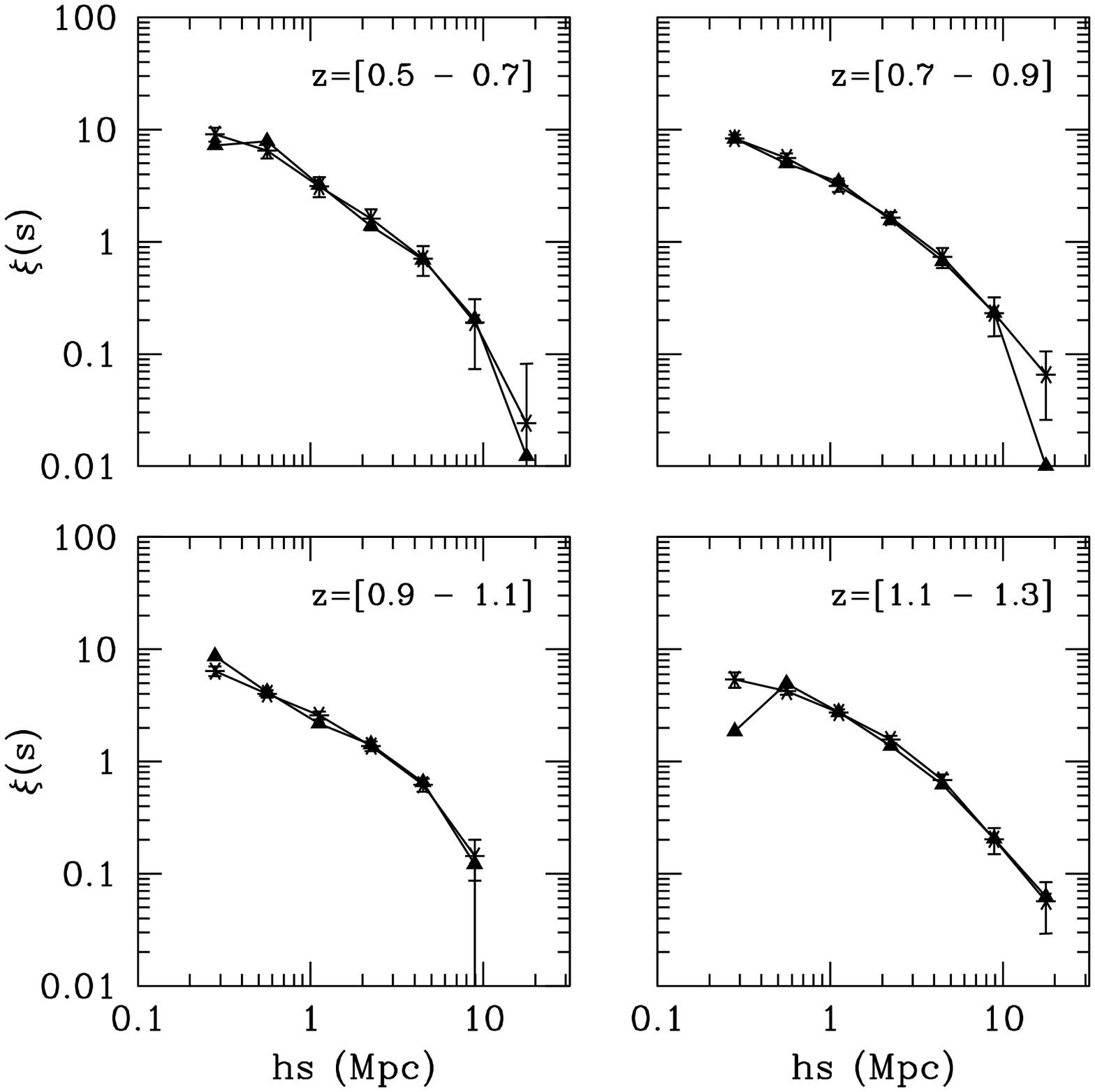}
   \caption{
Redshift-space two-point correlation function $\xi(s)$ for one 
mock VVDS-02h field, computed in four redshift bins.  The true
$\xi(s)$ computed for the whole parent sample (stars) is compared to that
measured from the ``observed'' sample, first without any correction
(open circles, left four panels) and then applying our correction
scheme (triangles, right four panels).
%{\bf The four panels in the left show the evolution of $\xi(s)$ with redshift, measured 
%in the original sample (stars) and, using the same "naive" method,
%after  the full observing strategy has been applied (open
%circles). The four right panels show the comparison of the same
%$\xi(s)$ measured from the parent  sample with the $\xi(s)$ measured
%in the "observed" catalog after the full correcting scheme is
%applied.  The 
Error bars are the $1\sigma$ ensemble {\it rms} among the 50
VVDS mock samples.}
% before (stars) and after (open circles)
%        the full observing strategy has been applied. 
% R! Despite the dramatic effect seen on $\omega(\theta)$, we note how
%$\xi(s)$ is in general not too sensitive to the observing biases.
          \label{xis_corr}  
    \end{figure*}

%______________________________________________
   \begin{figure*}
   \centering
   \includegraphics[width=8.5cm]{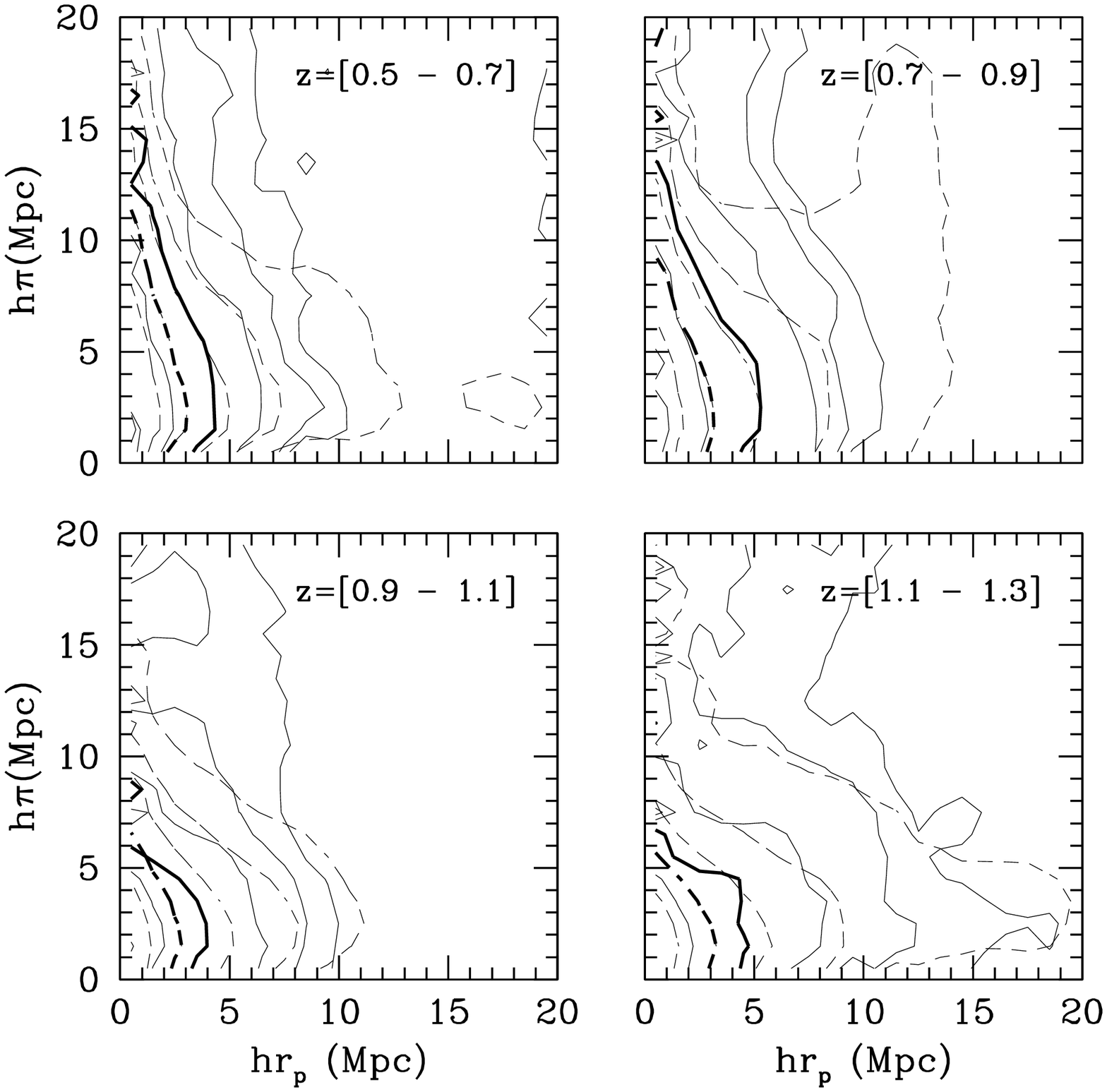}
   \includegraphics[width=8.5cm]{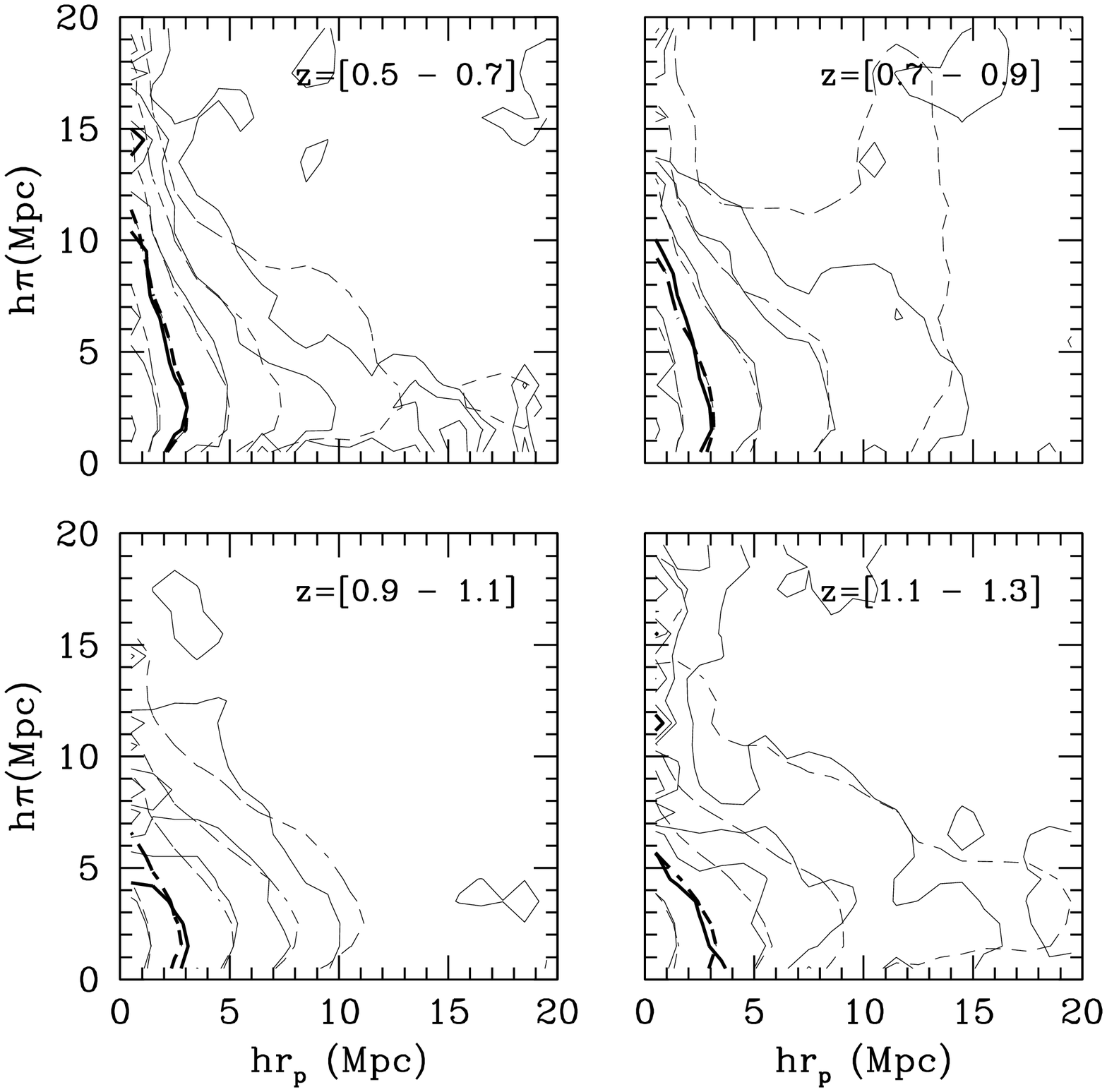}
   \caption{Same as Figure 9, but for the \xip correlation function.
%Impact of the observational selection function 
%on estimates of the redshift-space correlation
%function $\xi(r_p,\pi)$, performed on the same mock VVDS-02h field of previous
%figures.  
%The observed data have been smoothed with an isotropic
%Gaussian filter of width 3 $h^{-1}$ Mpc and 
The contours correspond to values for $\xi(r_p,\pi)$ of 0.4, 1 (bold),
2.0, 5.0.  Dashed lines refer to the complete mock sample, while solid
ones describe the sample after applying the VVDS selection
function. 
%{\bf The four left panels show the comparison of the 
%$\xi(r_p,\pi)$ "true" evolution and the one "measured" without 
%any correction. The impact of the observational biases on the 
%$\xi(r_p,\pi)$ is clearly visible here. The four right panels present 
%the comparison of the same values but when the full correcting scheme 
%has been applied to measure the $\xi(r_p,\pi)$ from the "observed" sample. }
 }
            \label{xip_corr} 
    \end{figure*}

   \begin{figure*}
   \centering
   \includegraphics[width=8.5cm]{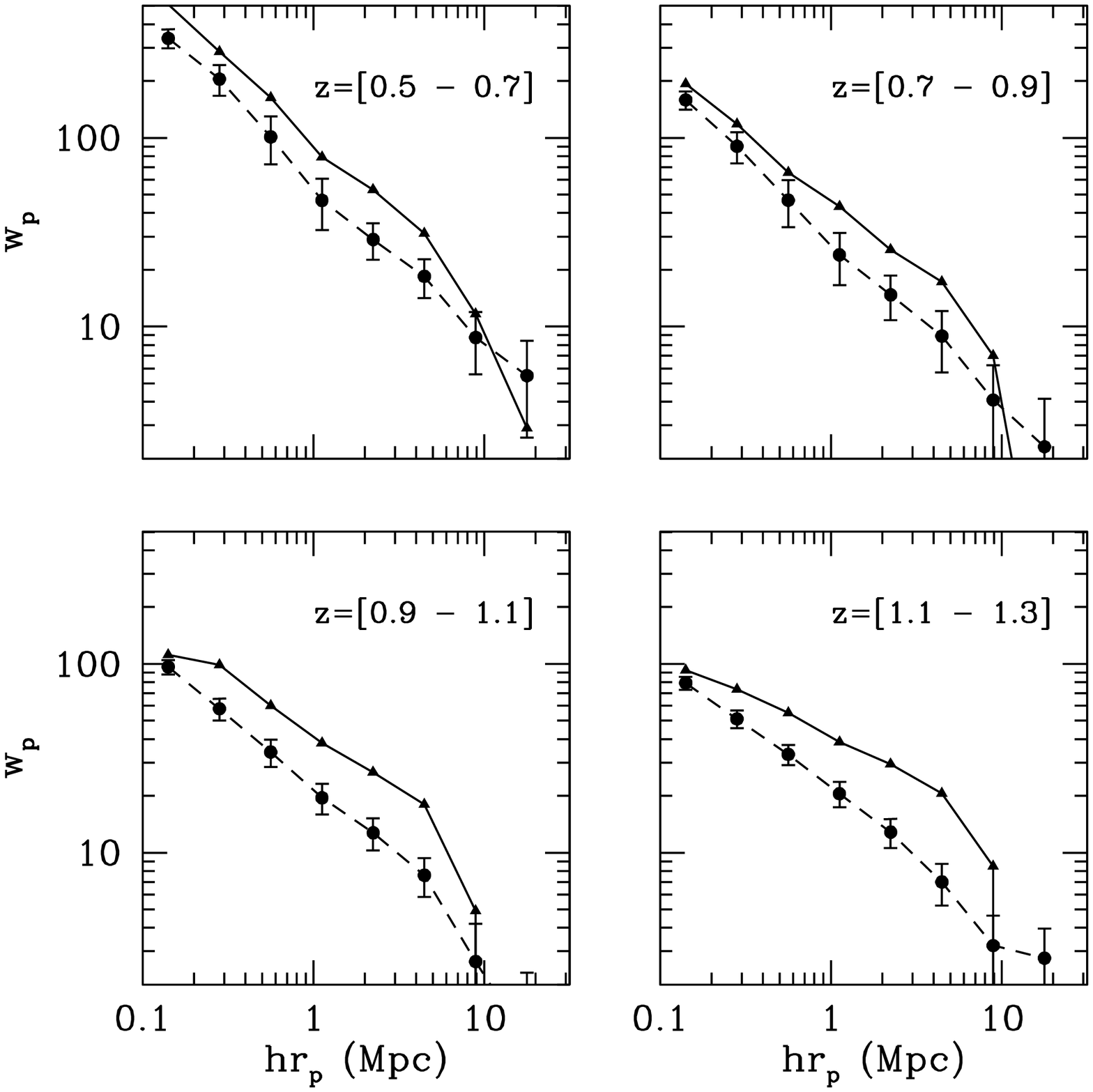}
   \includegraphics[width=8.5cm]{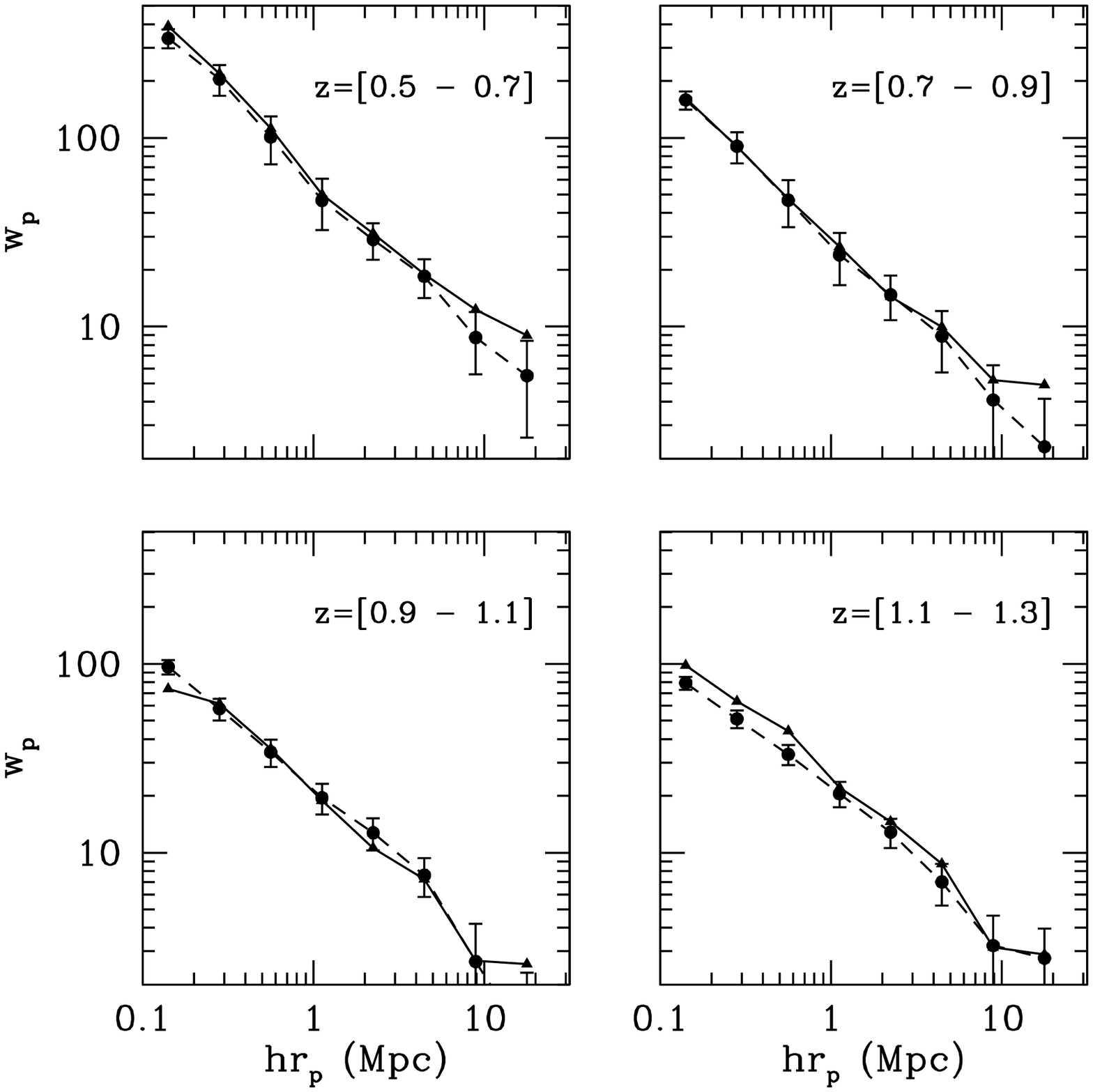}
   \caption{Same as Figures 9 and 10, but for the projected function
$w_p(r_p)$, 
% for one of the mock VVDS 02h fields, 
measured before (dashed line) and after (solid line) the full observing strategy has been
applied. 
%{\bf The four panels in the left show the comparison when no
% corrections are applied. In the four right panels the recovery of the
% evolution of $w_p(r_p)$ with the full correcting scheme presented in
% the section outlined in the section 4.2 is presented. Error bars are
% computed as described in the text.
This comparison shows that our method is able to properly recover
$w_p(r_p)$. We note, however, that, being closely related to the
angular function,
%all the correlation functions discussed here 
$w_p(r_p)$ remains the most sensitive among the 3D correlation
functions to the observational biases and the most difficult to recover properly in all $r_p$ bins.
%This, however, as we show in the figure \ref{evolution2}, does not affect significantly the measurement of $r_0$. 
} 
          \label{wp}
    \end{figure*}

%{\bf 5.1 and 5.2 moved HERE, MODIFIED and merged with 4.3 }

\subsection{
%Recovering Unbiased  Two-point Correlations - 
Correction scheme}
\label{correction}

The biases discussed so far involve introducing two types
of corrections which we discuss in detail in this section. 

1) {\it Global correction.} To account for the effects of uneven boundaries and varying sampling
rate we construct a random catalog, which consists of the same number 
of separately created pointings as the galaxy sample, thus reproducing 
the global ``exposure map'' (i.e. number of multiple passes 
over a given point of the sky) and the corresponding large-scale surface 
density variations of the galaxy redshift sample.  The holes and excised 
regions in the photometric sample are similarly taken into account by 
applying the same binary mask to the random sample.   
These first-order corrections account already for most of the observational 
biases.  When taken into account, they reduce most of the negative effects of 
the observing strategy on the correlation functions, in particular the global
overestimation of correlation functions (see Figures~\ref{xis_corr},~\ref{xip_corr},~\ref{wp}). 

%COMMENT - HERE WE CAN INTRODUCE AN ADDITIONAL, ``INTERMEDIATE'' PLOT 
%LIKE $xi(r_p,pi)$ AFTER THIS STEP? GGG: NO!

2) {\it Small scale correction.}
%{\it R!  Next, we need to correct for R!} 
What remains to be corrected is 
the slight bias introduced by the
slit-positioning software and the mechanical limitations (slit size,
closeness of slits and so forth).  We have seen that the
SSPOC selection is
not  an entirely random sampling of the actual 
angular distribution of objects, but rather a more homogeneous sub-set,
preferentially concentrated along specific rows.  This selection
affects primarily the small-scale values of the correlation function,
corresponding to the typical slit size: with only one spectroscopic
pass, pairs of galaxies with separation smaller than the slit size
will always have only one galaxy observed, and thus their contribution
to $\xi$ will be lost.  With repeated passes this problem is
alleviated, as the software chooses each time different objects (except for a
small number of objects observed twice for error checking purposes).
Using the full 2D information available from the parent photometric
catalog (that tells us how many galaxies on the sky have been missed in
the spectroscopic sample), we developed a weighting scheme that weighs
each targeted galaxy proportionally to its ``representativity'' in
terms of local angular pair density. 
%G2
%First, it will be useful to define a few numbers which we can measure 
%or estimate in a well-defined neighbourhood of a given $i$th galaxy 
%from a catalog. Let: \\

%G2
Let us therefore consider a circular region of radius $\theta_w$
around a galaxy $i$ located within a specific redshift slice $k$, and
define inside $\theta_w$ the following quantities:\\
$n_{gal}(i)$ -- the number of galaxies in the parent photometric catalog \\
$n_z(i)$ -- the number of galaxies with measured redshift \\
$n_{in}(i)$ -- the subset of these belonging to the same redshift
slice as the central galaxy\\
%and belonging to a given redshift slice\\
$n_{exp}(i)$ -- the number of
galaxies expected to belong to the same redshift slice, which can be
written as 
\begin{equation}
n_{exp}(i) = n_{in}(i)+n_{rem}(i)\,\,\,\,\,     ,
\end{equation}
with $n_{rem}$ being the fraction of unobserved neighboring galaxies in the parent
photometric catalog expected to belong to the same redshift slice.
This number can be written as
\begin{equation}
n_{rem}(i) = [n_{gal}(i) - n_z(i)]*P_{slice} \,\,\,\,   ,
\end{equation}
where $P_{slice}$ is the probability that a generic measured galaxy
belongs to that specific redshift slice. Here one can make the
reasonable assumption that the observed redshift distribution $N_z$ is
sufficiently well sampled as to provide, when averaged over a suitably
chosen radius $\theta_{A}$, an unbiased estimate of $P_{slice}$ for any $k^{th}$
slice as  
%the distribution of structures along the redshift direction. We can
%then use the observed $N(z)$, binned per redshift slices, to
%re-distribute the weight.   The probability of a galaxy to be
%within the $k$-th redshift slice is therefore
%
\begin{equation}
P_{slice} =
%\frac{N_{slice}}{n_z} = 
{N_{z,k} (<\theta_{A}) \over N_{z,total}(<\theta_{A})} \,\,\,\, .
\end{equation}
%
%where $\theta_{A}$ is a suitable radius, of the order of
%$20-30\arcmin$, within which the redshift distribution is estimated.
The choice of $\theta_{A}$ is clearly critical, as it has to be
large enough to allow a proper sampling of existing structures along
the line of sight (and thus minimize the noise introduced by the
weight), but also small enough not to dilute the effect of single
structures within one redshift slice.  In practice, given the
current size of the 02h field ($\sim 0.5$ square degrees), we
have obtained the best results using $\theta_{A}=30\arcmin$, which
encloses virtually the entire field. 
%For full consideration 
Note also that $n_{gal}(i)$, i.e. the number of galaxies in the parent
catalog, will be given by $n_{gal}(i) = n_{all}(i)*f_{gal}$, with
$f_{gal}$ being the probability that a randomly chosen object from the
photometric catalog is not a star but a galaxy and $n_{all}(i)$ 
- the number of all locally observed objects in this catalog.   
For the actual VVDS 02h field, this
probability has been estimated to be $f_{gal}=0.92$. 

%GG2 Since 
The construction of the actual weight to recover the loss of
small-scale pairs produced essentially by the proximity bias is not
unequivocal. After several experiments with weighting by local
densities (of expected vs. observed spectra), we obtained the best
results weighting by pairs.
The two-point correlation function being a pair-weighted statistic,
%is based on the pair-counting, 
we constructed our weight $w(i)$ for a given galaxy $i$
%may be constructed from 
from the ratio of the expected to the measured number of pairs within
$\theta_w$.  Specifically, if one wants the local angular pair density
to be conserved, each pair should be counted as:
\begin{equation}
w(i)*w(j) = \frac{n_{exp}(i)*(n_{exp}(j)-1)}{n_{in}(i)*(n_{in}(j)-1)}\,\,\,\,   .
\end{equation}
And, consequently, a single object is assigned a weight  
\begin{equation}
w(i)=
\sqrt{\frac{n_{exp}(i)*(n_{exp}(i)-1)}{n_{in}(i)*(n_{in}(i)-1)}}\,\,\,\,        .
\end{equation}
To define the optimal angular size $\theta_w$ defining the
``neighborhood'' of a galaxy,  
%Now remains the problem of defining the best "local" scale 
%to compute the weight. All local densities are computed in the neighborhood
%of the given galaxy, inside a circle of a radius $\theta_w$.  To find
%the optimal choice of $\theta_w$, 
we experimented with different values in the range 5$\arcsec$ to 1$\arcmin$.  Not
surprisingly, the best correction is obtained for $\theta_w$ in the
range $30 - 45\arcsec$, which is comparable to the length of the
VIMOS spectra as projected on the sky.   In all computations presented
here, we adopted the value $\theta_w = 40''$.  

The following sections will present the results of extensive tests of
this correction scheme,
%using
based on the \galics{} mock VVDS surveys.

\subsection{Application to redshift-space correlations}

We have applied the manipulations presented in the previous section to our
mock VVDS 02h
%fields
surveys and compared the results to those obtained from the
whole 1$\deg \times 1 \deg$ mock 
%survey
fields.  
%GGG
%When {\bf extending the analysis} from 2D to 3D 
%with statistical analysis,
%{\it  things change significantly. }
%{\bf the effect of the observing biases becomes less critical but
%remains significant. }
The results are shown in Figures~\ref{xis_corr},~\ref{xip_corr} and
\ref{wp} for the same mock sample
%(complete and ``observed'') 
used for measuring $\omega(\theta)$  (Figure~\ref{ang1}), split into 4
redshift bins.  In each of these figures, comparison of the the
four left to the four right panels demonstrates the
effect of the overall correction.
%of Figure show the comparison of $\xi(s)$ true and measured from the "observed" sample but }
%and not corrected for the observing biases. 
In general, 
%immediately that once 
in redshift space the effect of the observational biases  
%introduced by the observing strategy becomes 
is much less severe, being diluted by the unaffected clustering measured along 
the line of sight.  Still, we see how a proper estimate 
%of $\xi(s)$ 
does require a correction. 

Looking at $\xi(s)$ (Figure~\ref{xis_corr}), we see that the correction introduced by our scheme is 
in general very good. 
% as shown in {\bf four right panels of} figure ~\ref{xis_corr}. 
The full bi-dimensional correlation function $\xi(r_p,\pi)$
(Figure~\ref{xip_corr}) shows the effect in more detail, indicating  
%shows the classical representation of $\xi(r_p,\pi)$ as a contour
%plot.  Here, {\bf in the four left panels}, before any corrections
%are applied, the impact  of the observational {\bf survey} selection
%function on the shape of $\xi(r_p,\pi)$ contours is evident.   
%We can also see 
also that the impact of the angular bias on spatial correlations 
depends on redshift. This is to be expected, given that a fake 
inhomogeneity at a given angular scale affects larger spatial scales 
at larger redshifts.
%, thus introducing a spurious clustering signal on large scales.
However, as seen from the four right panels
%figure~\ref{xip_corr}, 
%most of these problems are solved after the correcting scheme is
%applied. 
the bulk of the problem is corrected by our technique.

Finally, the 
%computed
corresponding projected function,
%After the full correcting scheme is applied, 
%the projected function 
\wp, which is the one that will be fitted to estimate the real-space correlation 
length and slope, 
(Le F\`evre et al. 2005), does not show 
any significant systematic effect, nor scale-dependent bias (see also
\S~\ref{galics_evol} below), if one excludes a residual effect
in the highest-redshift bin (which may be specific of the mock sample
used).
% any more. 
%GGG
\subsection{Accuracy in recovering $r_0$ and $\gamma$} 
\label{galics_evol}

%GGG I DON'T LIKE THIS
%We identify as the ``best'' pair $(r_0,\gamma)$ the values for which 
%the $\chi^2$ is minimal. The $\chi^2$ map is computed as a
%function of $r_0$ and $\gamma$ in order to produce error contours
%around the best solution. {\bf BM: $\chi^2$ -> CONFIDENCE LEVELS}

\begin{figure}
   \centering
   \includegraphics[width=9cm]{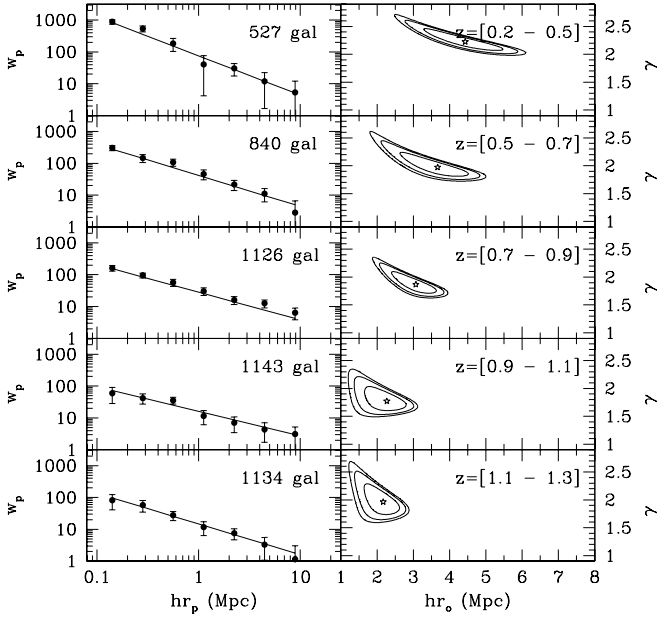}
   \caption{Evolution of the projected function \wp (left column) and
the corresponding best-fit parameters of \xir, $r_0$ and $\gamma$ (right column), 
%GGG OBVIOUS: as a function of redshift 
as seen in one of the VVDS mock surveys. 
%Left panels: $w_p$ computed in five redshift bins. 
Error bars are computed as explained in the text, while 
%Right panels: the $(\gamma,r_0)$ {\bf pair} %couple with the 
error contours on the fit parameters are obtained taking into account
the full covariance matrix.
%correlation between the bins. 
%The contours are plotted at the
The $68.3\%$, $90\%$ and $95.4\%$ joint confidence levels are defined as in Numerical Recipes 
(\cite{NumRecip}, chapter 15.6) in terms of the corresponding likelihood intervals that
we obtain from our fitting procedure (see \S~\ref{fit}).} 
        \label{evolution}
   \end{figure}

\begin{figure}
   \centering
   \includegraphics[width=8.5cm]{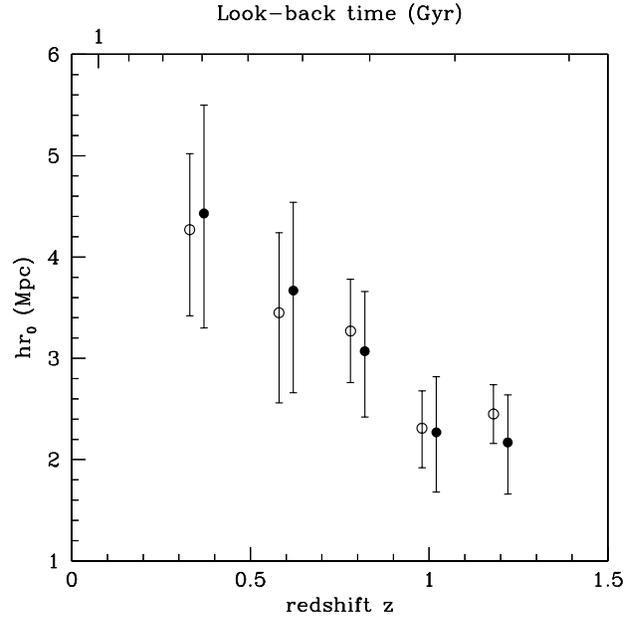}
   \caption{Evolution of $r_0$ in a VVDS mock survey (filled circles),
   compared to that of its parent catalog (open circles).  Error
   bars are as explained in the text. The "true" and "measured"
   values of $r_0$ are very consistent within the error bars,
   providing an {\it internal} proof of the quality of our
   correction scheme. }
\label{evolution2}
\end{figure}

Let us now evaluate more quantitatively how well the weighting
scheme 
%described in theprevious section 
is able to recover the correct values of the two
parameters of \xir, $r_0$ and $\gamma$.   
%As an example of this process,we present in 
Figure \ref{evolution} plots the projected correlation function
$w_p$, computed for one of the VVDS mock cones, together with 
the measured best fit values of $r_0$ and $\gamma$.
%As expected, the correlation length $r_0$ in the
%simulations decreases with the redshift as the 
%underlying large scale structures follow the 
%%evolution of the (dark) matter.  
The error contours are estimated from the variance of the 50 mock surveys as
described previously and their size 
%of the error contours in the different bins 
depends mainly on the number of galaxies within each bin.
%observed in these bins with our observing strategy.  
Figure~\ref{evolution2} shows that the evolution of clustering we ``observe'' in
this specific simulated VVDS cone agrees quite well with its parent
sample.
%well with the clustering predicted by \galics simulations in general, as shown in Figure. 
%GGG THIS DOES NOT DEMONSTRATE IT The result demonstrates that it is
%indeed possible to measure the  ``true'' clustering evolution from the VVDS data.

\begin{figure}
   \centering
    \includegraphics[width=8.5cm]{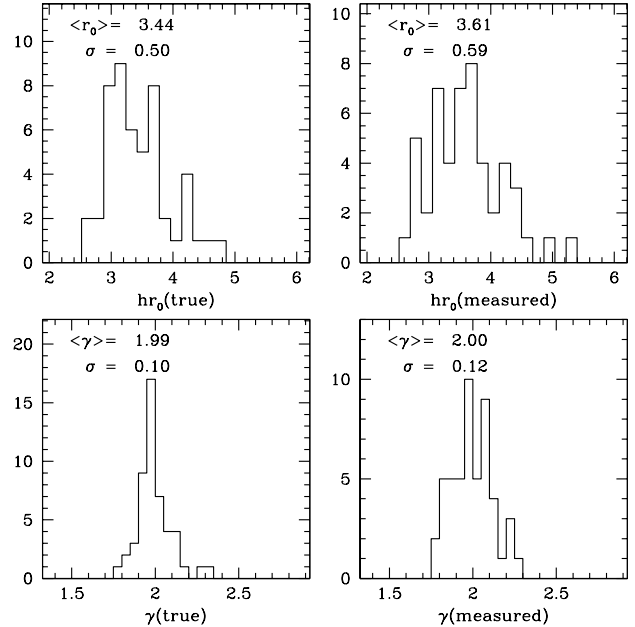}
      \caption{Histograms of the measurements of 
$r_0$ and $\gamma$ in the redshift bin [0.5 - 0.7]
(chosen as a representative case), among the $50$ mock catalogs, for the full 
cones (left column) and for the observed samples (right column), where
the full weighting scheme has been applied.
The ensemble averaged values of $r_0$ and $\gamma$  are indicated in
each panel, together with their {\it rms} error.       }
         \label{rgafter}
  \end{figure}

Of course, due to cosmic variance, the values of $r_0$ and $\gamma$
differ between different simulated cones.  Figure ~\ref{rgafter} shows
the spread of these parameters among all the $50$ mock VVDS surveys
and their parent catalogs, for a representative redshift bin ($z=[0.5-0.7]$).  This
behavior is similarly seen in the other redshift bins, indicating an
increased spread in the parameter estimates in the ``observed''
catalogs, an effect easily explained in terms of the smaller number of objects.
%$\gamma$ and $r_0$ measured from the mock VVDS surveys and their parent samples.
%and of the order of 10\% in case
%of $\gamma$ (while the measurement of $r_0$ and $\gamma$ are, of
%course, not independent).  
Figure~\ref{evolution} and
Figure~\ref{rgafter} also indicate that 
at the end of our correction process 
%there is still a  tendency to {\bf overestimate} slightly but
%systematically $r_0$.   The mean correlation length $r_0$ after applying
%our correcting scheme is on average 5\%  {\bf higher} than the ``true'' correlation 
%length in the mock samples.  
any possible systematic effect is reduced to less than 5\%, a
value 
%This difference, however, is 
always significantly smaller than the %larger error bars coming from 
uncertainty due to cosmic variance which is %are 
of the order of 15--20\%. 
%GGG REDUNDANT: As figure~\ref{evolution2} shows, our measured values are always
%consistent with the respective values in the parent catalogs, within the measurement errors.

%these 2 figures were made with GalICS cone022

\subsection{Tests of VVDS observing strategy}

In this Section we want to 
%check 
discuss from a more general perspective (i.e.
not limited to the current status and lay-out of the 02h field) how
%much the measured value and 
the accuracy of correlation measurements can
%functions 
depend on
the number of multiple spectroscopic pointings (``passes'') that are
dedicated to a specific area.  In other words: are multiple passes
increasing --- as expected --- the accuracy of correlation function
measurements, not only thanks to the increased statistics, but also
because of the improved sampling of the clustering process?  And how is
our correcting scheme performing when handling a very sparse (one
pass) or a more densely sampled area?  This is clearly an
interesting question for the future development of the VVDS, 
or other surveys, as these
tests can indicate what strategy could be more efficient.
%GGG in terms of measuring correlation functions.  
%GGG REDUNDANT 
One would like to estimate the fraction of galaxies 
%it is 
necessary
%to observe spectroscopically 
to recover the correlation signal
to a certain level of accuracy. This,
% compared to the total number of targets available from the deep photometry,  or, 
translated to the VVDS, implies determining how many spectroscopic ``passes''
with VIMOS are necessary. Note that the answer is
not trivial, since multiple pointings over the same area are usually
dithered (i.e. shifted by an amount at least larger than the central
``cross'', i.e. 2\arcmin), and thus a larger number of passes over
the same area, while improving the sampling, introduces also a more
complex mean density pattern, as explained in section 2.1.

\begin{figure}
   \centering
   \includegraphics[width=8.5cm]{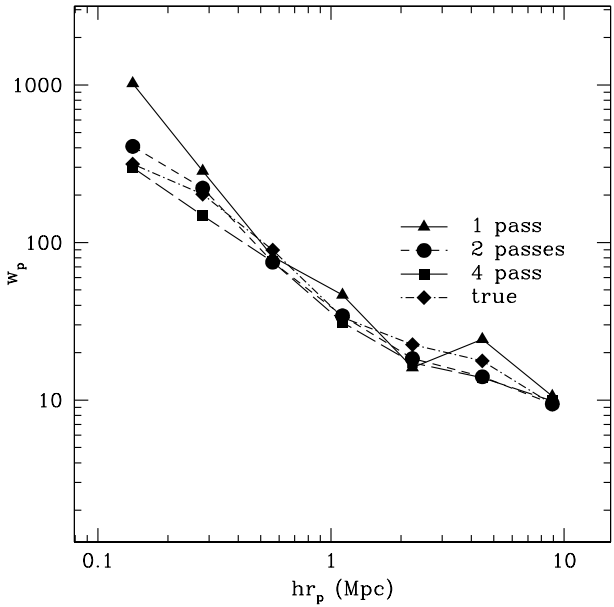}
      \caption{Measured $w_p(r_p)$ in the case of different number
        of passes over the same field.  When the field is observed
        only once we 
%have an obvious (however small) lack of power 
are clearly not able to properly recover properly $w_p(r_p)$
        on the smallest scale. When we observe the field more times 
        the recovery is much better also on the small scales. 
%{\it In all cases, however, there remains a slight bias toward smaller values of $w_p(r_p)$.
}
         \label{f10wp}
   \end{figure}

\begin{figure}
   \centering
   \includegraphics[width=8.5cm]{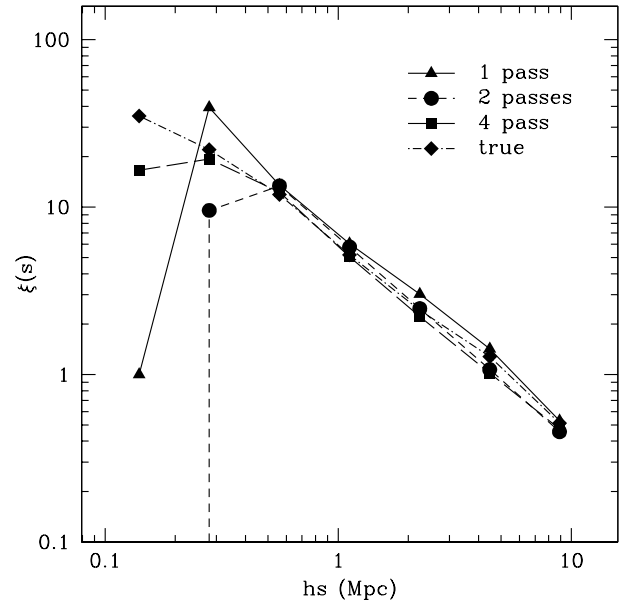}
     \caption{Measurements of $\xi(s)$ for a different number of observing
``visits'' over the same field.
        }
         \label{f10xis}
   \end{figure}

Tests have been performed creating a grid of six pointings, spaced
with the same step as the real VVDS ones in the VVDS-10h field. The second
pass was then arranged over a grid shifted by %around 
$2\arcmin$ in right ascension and declination. The pointings of  both passes 
have then been "observed" once again with a different selection of objects
for spectroscopy. At the end (maximum coverage), this
resulted in an area of $0.3624$ square degrees, mostly uniformly covered 
but with small patches of sky that were observed either three, %or 
two or one times or remained unobserved.  
The results for \wp and \xis are shown in Figures~\ref{f10wp} and
\ref{f10xis}, respectively.  

The %corrected 
projected correlation function $w_p$ is fairly well
recovered almost independently of the sampling density. 
For a single pass, power is 
%lacking 
not recovered properly at %smallest
scales below $0.6$ h$^{-1}$Mpc, since there is in practice no pair (even biased) to be
``corrected'' in a proper statistical way by our scheme.  
%In other words, clustering on the smallest scales (smaller than the 
%slit size) is simply beyond recovery. 

The case of \xis (Figure~\ref{f10xis}) shows even more clearly the
difficulty of recovering very small scale pairs with only one pass:
in this case, there is an intrinsic low-scale limitation (complete
lack of pairs), which cannot be fully overcome by the correcting
scheme.  The figure shows, for example, that while a linear bin
between 0 and 1 $\hmpc$ is already sufficient to recover the
correct clustering amplitude even with one pass, smaller
logarithmic bins below 1 $\hmpc$ are inadequate and suffer from the lack
of measured pairs.

We conclude that even in the fields that were observed only with one 
spectroscopic observation, sampling about 15\% of the photometric
targets down to $I_{AB}=24$, 
the two-point correlation function can be measured quite well
for separations $1 \leq r \leq 10 h^{-1}$ Mpc.
The results confirm, however, that observing fields four times,
sampling about 40\% of the population
as in the deep part of the VVDS, provides the possibility of
more precise measurements on %small 
scales %below 
down to 0.1 $h^{-1}$ Mpc.

\section{Summary and conclusions}

One of the key goals of the VVDS survey is to measure the evolution of
the galaxy clustering from the present epoch up to $z\sim 2$ and
larger. To study in detail the error budget of $\xi(r)$ measurements in
the VVDS survey, we have generated a set of mock catalogs using the
GalICS model of semi-analytic galaxy formation. The geometry of the
VVDS survey on the sky is complex due to the observing strategy. The
resulting selection function substantially affects the angular correlation
properties of the clustering of the observed galaxies.  We demonstrate
that the correlation observed in redshift space is much less
affected and that the
bias introduced by the observing strategy can be largely removed using
the correcting scheme we propose in this paper.

We conclude that, for the first epoch VVDS data, we can expect
to measure $\xi(s)$ and $w_p(r_p)$ to better than 10\% on scales 
$1 \leq r \leq 10$ h$^{-1}$ Mpc, and better than 30\% below 1 h$^{-1}$ Mpc. 
%The projected correlation
%function $\w_p(r_p)$, is usually recovered
%with an accuracy better than 10\%.  
Results obtained from the
GalICS simulations indicate that the two-point correlation functions
computed from the First Epoch VVDS should suffer only from a modest cosmic
variance of $\simeq15-20$\%. These results suggest
that after the final selection of objects for spectroscopy the variance
becomes twice as large as the variance of the underlying parent galaxy 
field in the same area. We expect,
in each redshift slice $\Delta z\simeq0.2$ in the 
redshift range z=[0.2,2.1], to measure $r_0$ and $\gamma$
with an accuracy better than $15-20\%$. We show
%GGG: LET'S BE POSITIVE!
that any residual systematic effect in the measurements of $r_0$
and $\gamma$ is 
%after  correcting scheme remain systematically 
%{\it smaller}
below
%, due to the observing constraints, than the correlation length
%of the parent sample by about 
5\%, i.e. a value 
%This offset, however, 
much smaller than the cosmic errors.

The actual measured clustering properties of galaxies in the VVDS survey, 
using the framework outlined in this paper, are presented in
\cite{CF_OLF} and in forthcoming papers.

\begin{acknowledgements}
We thank the \galics{} group for access to their simulations, S. Colombi
for providing a first set of mock n-body samples early in the
development of this work and for useful discussions.  This research
has been developed within the framework of the VVDS consortium and has
been partially supported by the CNRS-INSU and its Programme National
de Cosmologie (France), and by Italian Research Ministry (MIUR) grants
COFIN2000 (MM02037133) and COFIN2003 (num.2003020150).
\end{acknowledgements}

\end{document}